\documentclass[aps,prb,notitlepage,twocolumn]{revtex4-1}%
\usepackage{hyperref}
\usepackage{natbib}
\usepackage{amsfonts}
\usepackage{amsmath}
\usepackage{amssymb}
\usepackage[pdftex]{graphicx}%
\begin{document}
\title{Magnetotransport of multiple-band nearly-antiferromagnetic metals due to
``hot-spot'' scattering}
\author{A. E. Koshelev}
\affiliation{Materials Science Division, Argonne National Laboratory, Argonne, Illinois 60439}
\date{\today }

\begin{abstract}
Multiple-band electronic structure and proximity to antiferromagnetic (AF)
instability are the key properties of iron-based superconductors. 
We explore the influence of scattering by the AF spin fluctuations on transport of
multiple-band metals above the magnetic transition. A salient feature of scattering on the
AF fluctuations is that it is strongly enhanced at the Fermi surface locations where the
nesting is perfect (``hot spots'' or ``hot lines''). We review derivation of the collision
integral for the Boltzmann equation due to AF-fluctuations scattering. In the
paramagnetic state, the enhanced scattering rate near the hot lines leads to anomalous
behavior of electronic transport in magnetic field. We explore this behavior by
analytically solving the Boltzmann transport equation with approximate transition rates. This
approach accounts for return scattering events and is more accurate than the
relaxation-time approximation. The magnetic-field dependences are characterized by two
very different field scales, the lower scale is set by the hot-spot width and the
higher scale is set by the total scattering amplitude. A conventional magnetotransport
behavior is limited to magnetic fields below the lower scale. In the wide range in between
these two scales the longitudinal conductivity has linear dependence on the magnetic field
and the Hall conductivity has quadratic dependence. The linear dependence of the diagonal
component reflects growth of the Fermi-surface area affected by the hot spots proportional to
the magnetic field. We discuss applicability of this theoretical framework for describing
of anomalous magnetotransport properties in different iron pnictides and chalcogenides in the
paramagnetic state.
\end{abstract}
\maketitle

\section{Introduction}

\begin{figure}[ptb]
\includegraphics[width=2.5in]{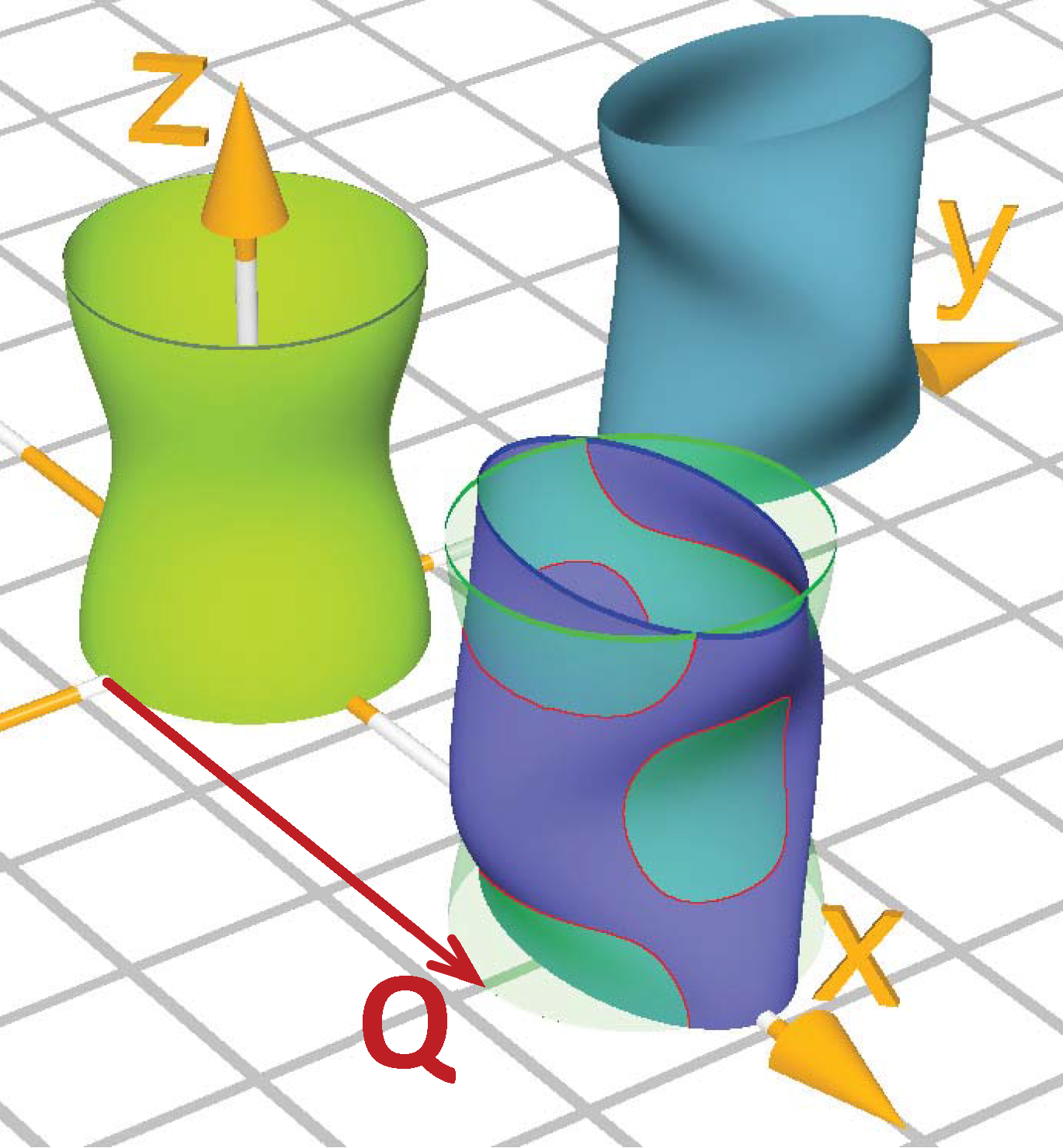}
\caption{Schematic Fermi surface typical for iron pnictides showing only one hole band and
	two electron bands. Intersection between the hole Fermi surface displaced by the AF
	ordering wave vector $\mathbf{Q}$ and the electron Fermi surface marks the hot lines.
	Scattering by the AF fluctuations is enhanced at these lines. }%
\label{Fig-FermiSurf3D}%
\end{figure}
Rich normal-state properties of iron-based high-temperature superconductors are caused by proximity
to antiferromagnetic (AF) transition and multiple-band electronic
structure.\cite{PaglioneNatPhys10,StewartRMP11,ShibauchiARCMP14,HosonoPhysC15} AF fluctuations play
important role in these materials and it is likely that superconductivity is mediated by these
fluctuations. One can expect also that the spin fluctuations scatter quasiparticles in normal state
and therefore influence transport properties. In particular, linear temperature dependence of
resistivity near optimal doping \cite{GoochPhysRevB09,KasaharaPhysRevB10} has been attributed to AF
fluctuations near the quantum critical point. Such spin-fluctuations scattering is the strongest
when momentum transfers are close to the AF instability vector $\mathbf{Q}$. As a consequence,
scattering rate is strongly enhanced near so-called ``hot lines'' (or ``hot spots'' in quasi
two-dimensional case), corresponding to ideal-nesting conditions for the vector $\mathbf{Q}$, see
Fig.\ \ref{Fig-FermiSurf3D}. The concept of hot spots has been introduced for cuprate
high-temperature superconductors and their role in the transport properties has been considered in
several theoretical papers
\cite{HlubRicePhysRevB94,StojkovicPinesPhysRevB97,LohneysenRevModPhys.79.1015,KontaniRPP08}. For
iron pnictides, the effects due to hot-spot scattering also has been discussed
\cite{FernandesPhysRevLett11,BreitkreizPhysRevB14,BreitkreizAnisPhysRevB14}. 
In particular, the resistivity anisotropy induced by the orthorhombic deformation has been
considered in Refs.\ \cite{FernandesPhysRevLett11,BreitkreizAnisPhysRevB14}. It was demonstrated
that the hot-spot scattering mechanism provides a consistent description of the experimental
anisotropy dependences on temperature and doping\cite{BlombergNComm13}. %
In the related work \cite{BreitkreizPhysRevB13} the effects caused by  interband scattering by AF
fluctuations in metals with multiple isotropic bands have been investigated. Even though the hot
lines are absent in this situation, it was demonstrated, nevertheless, that strong and anisotropic
interband scattering leads to anomalous transport properties which are not described by the simple
relaxation-time approximation. 

The narrow hot lines do not strongly change conductivity which is determined by the
average scattering time and therefore regular ``cold'' regions with weak scattering
dominate.\cite{HlubRicePhysRevB94} However, the hot lines give the anomalous behavior of
the conductivity in magnetic field \cite{RoschPhysRevB00} including possible extended
range of linear decrease with field. Such anomalies appear because the regions on the
Fermi surface influenced by the hot lines grow with magnetic field. Similar mechanism also
leads to unusual magnetic-field dependence of the Hall conductivity.

Electronic transport in the magnetic field has been investigated in detail
practically for all families of iron-based superconductors spanning wide
ranges of dopings
\cite{RullierAlbenquePhysRevLett09,RullierAlbenquePhysRevB10,TsukadaPhysRevB.81.054515,ShenPhysRevB11,OhgushiPhysRevB12,EomPhysRevB12,SunPhysRevB14,LiPhysRevB.90.024512,AnalytisNPhys14,MoseleyPhysRevB15}. 
The high-field magnetotransport in the paramagnetic state do exhibit several
anomalous features such as linear magnetoresistance in Fe$_{1+y}$Te$_{0.6}
$Se$_{0.4}$ \cite{SunPhysRevB14}, nonquadratic magnetoresistance in
optimally-doped Ba[As$_{1-x}$P$_{x}$]$_{2}$Fe$_{2}$
\cite{AnalytisNPhys14,WelpUnpubl} and FeSe\cite{WatsonPhysRevLett15}, as well
as the strongly nonlinear Hall resistance in FeTe$_{0.5}$Se$_{0.5}
$\cite{TsukadaPhysRevB.81.054515}, Ba$_{0.5}$K$_{0.5}$As$_{2}$Fe$_{2}$
\cite{LiPhysRevB.90.024512}, FeSe \cite{WatsonPhysRevLett15}, and
Ba[As$_{1-x}$P$_{x}$]$_{2}$Fe$_{2}$\cite{WelpUnpubl}.
These effects are likely to be caused by the hot-spot scattering due to the AF
fluctuations. However, strong anisotropy of the spin-fluctuation scattering is
frequently ignored and transport properties of the iron pnictides are
interpreted using more conventional multiple-band Fermi-liquid theory
\cite{RullierAlbenquePhysRevLett09,RullierAlbenquePhysRevB10,ShenPhysRevB11,OhgushiPhysRevB12,
WatsonPhysRevLett15} assuming that all scattering channels can be fully
characterized by the band-dependent scattering rates. 

Motivated by a clear relevance of the hot-spot mechanism for the iron pnictides and chalcogenides, we
investigate in this paper transport properties of nearly-antiferromagnetic multiple-band metals. We
consider in detail derivation of the collision integral for the Boltzmann equation and quasiparticle
lifetime due to scattering by spin fluctuations. This allows us to relate the shape and strength of
the hot-spot scattering rate with the microscopic parameters of the system. We proceed with
analytical solution of the Boltzmann transport equation in the magnetic field using approximate
transition rates which reproduce correctly physics of the hot-spot scattering. Our approach fully
accounts for the return scattering events and therefore it is more accurate than the widely-used
relaxation-time approximation. Based on the derived distribution functions, we compute the
magnetic-field dependences of the longitudinal and Hall conductivities. These dependences are
characterized by the two very different magnetic-field scales, the lower scale is set by the width of
the hot spots and the higher scale is set by the total scattering amplitude. A conventional
magnetotransport behavior is limited to the magnetic fields below the lower scale. In the wide field
range in between these two scales the longitudinal conductivity has linear dependence on the
magnetic field and the Hall conductivity has quadratic dependence. The linear dependence of the
diagonal component reflects growth of the Fermi-surface area affected by hot spots proportional to
the magnetic field. In the intermediate range the conductivity components are almost independent of
the hot-spot parameters.

A somewhat similar behavior of magnetotransport is also realized in the antiferromagnetic
state due to the Fermi surface reconstruction near the nesting points caused by 
opening of the antiferromagnetic gap\cite{FentonPRL05,LinPRB05,KoshelevPRB13}. The
reconstructed Fermi surface acquires turning points at which the Fermi velocity changes
abruptly. As a consequence, the longitudinal conductivity has linear dependence on the
magnetic field \cite{FentonPRL05,KoshelevPRB13} and the Hall component has quadratic
dependence \cite{LinPRB05} above the field scale set by the antiferromagnetic gap and
scattering rate. In contrast to the hot-spot mechanism, for isolated turning points there
is no higher magnetic field scale limiting this behavior from above.

This paper is organized as follows. In Sec.\ \ref{Sec:Model} we introduce the microscopic model
describing two-band metal interacting with AF fluctuations. Based on this model, we present
derivation of the collision integral in the Boltzmann equation in Sec.\ \ref{Sec:CollInt}. We
analytically solve the Boltzmann equation using approximate scattering rates in Sec.\
\ref{Sec:SolBoltz}. In Sec.\ \ref{Sec:CondZeroH} we compute conductivity in zero magnetic field. The
conductivity components in magnetic field are considered in Sec.\ \ref{Sec:CondMagnField}
(longitudinal conductivity in subsection \ref{Sec:LongCond} and Hall conductivity in subsection
\ref{Sec:HallCond}). In Section \ref{Sec-MagFieldDep} we illustrate typical magnetic-field
dependences of the conductivity components for a simple four-band model with two electron bands and
two identical hole bands.

\section{Microscopic Model}
\label{Sec:Model}

For electronic band structure of iron-pnictides, the fluctuating AF magnetization mixes two bands,
electron and hole. This means that for the treatment of an isolated hot line, it is sufficient to
consider only a pair of interacting bands described by the following Hamiltonian
\begin{equation}
\mathcal{H}=\mathcal{H}_{0}+\mathcal{H}_{\mathrm{AF}},\label{Hamilt}%
\end{equation}
where the free-electron part is composed of the electron and hole
contributions,
\begin{equation}
\mathcal{H}_{0}=\sum_{\mathbf{p},\sigma}\left(  \xi_{1,\mathbf{p}
}c_{\mathbf{p}\sigma}^{\dagger}c_{\mathbf{p}\sigma}+\xi_{2,\mathbf{p}}
d_{\mathbf{p}\sigma}^{\dagger}d_{\mathbf{p}\sigma}\right)  .\label{HamFree}%
\end{equation}
A particular shape of spectrum is not important for further consideration. In
the electron part $\xi_{1,\mathbf{p}}$ the momentum $\mathbf{p}$ is measured
with respect to the lattice wave vector $\mathbf{Q}$ at which the AF ordering
takes place. The Fermi surfaces are determined by $\xi_{s,\mathbf{p}}=\mu$,
where $\mu$ is the chemical potential. The hole Fermi surface and displaced
electron Fermi surface cross along the hot lines, i.e., where $\xi
_{1,\mathbf{p}}=\xi_{2,\mathbf{p}}$.

The antiferromagnetic part of the Hamiltonian is given by
\begin{equation}
\mathcal{H}_{\mathrm{AF}}\!=\!-\frac{g}{2}\!\sum_{\mathbf{p},\mathbf{p}
^{\prime},j,\alpha,\beta}\!(M_{j,\mathbf{q}}\sigma_{\alpha\beta}
^{j}c_{\mathbf{p}\alpha}^{\dagger}d_{\mathbf{p}^{\prime}\beta}
\!+\!M_{j,-\mathbf{q}}\sigma_{\beta\alpha}^{j}d_{\mathbf{p}^{\prime}\beta
}^{\dagger}c_{\mathbf{p}\alpha}),\label{HamAF}%
\end{equation}
where $M_{j,\mathbf{q}}$ are the magnetization components, $j=(x,y,z)$,
$\mathbf{q}\!=\!\mathbf{p}\!-\!\mathbf{p}^{\prime}$ is the shift of the wave
vector with respect to the AF-ordering vector $\mathbf{Q}$, and $\sigma
_{\alpha\beta}^{j}$ are the Pauli matrices.

In paramagnetic state $\mathbf{M}_{\mathbf{q}}\equiv\tilde{\mathbf{M}
}_{\mathbf{q}}(t)$ is the fluctuating magnetization which, in particular,
scatters the carriers between the bands. According to the fluctuation-dissipation
theorem, the amplitude of fluctuating magnetization $\left\langle \left\vert
\tilde{M}_{j}(\mathbf{q},\omega)\right\vert ^{2}\right\rangle $ is connected
with the magnetic susceptibility $\chi_{j}(\mathbf{q},\omega)$ as
\begin{equation}
\left\langle \left\vert \tilde{M}_{j}(\mathbf{q},\omega)\right\vert
^{2}\right\rangle =\frac{2 T}{\omega}\operatorname{Im} \left[  \chi
_{j}(\mathbf{q},\omega)\right] \label{FDT}%
\end{equation}
for $T\gg\omega$ (classical limit)\footnote{We use system of units with
$k_{B}=1$ and $\hbar=1$ throughout the paper.}. The commonly used form of the
susceptibility
\begin{equation}
\chi_{j}(\mathbf{q},\omega) =\frac{1}{-i\gamma\omega+\alpha_{j}+\eta_{i}
q_{i}^{2}}\label{GaussSuscept}%
\end{equation}
is valid for weak Gaussian magnetic fluctuations. In this case
\begin{equation}
\left\langle \left\vert \tilde{M}_{j}(\mathbf{q},\omega)\right\vert
^{2}\right\rangle =\frac{2\gamma T}{\gamma^{2}\omega^{2}+
	\left(\alpha_{j}+\eta_{i}q_{i}^{2}\right)^{2}}.
\label{GaussFluctMagn}%
\end{equation}
The parameters $\alpha_{j}$ with $j=x,y,z$ characterize proximity to the
magnetic transition temperature $T_{S}$. For continuous phase transition at
least one of these parameters vanish at $T_{S}$. We mention that for
continuous phase transitions the simple shape of the susceptibility
(\ref{GaussSuscept}) is not valid in the vicinity of the transition point, in
the regime of strong critical fluctuations.

\section{Collision integral in Boltzmann equation and quasiparticle lifetime}
\label{Sec:CollInt}

Scattering of carriers are fully characterized by the collision integral in
the Boltzmann equation \cite{ZimanBook,BlattBook}. The collision integral for
scattering on the AF antiferromagnetic fluctuations was derived in
Refs.\ \onlinecite{HlubRicePhysRevB94} and \onlinecite{RoschPhysRevB00} for
the two-dimensional and three-dimensional cases correspondingly. In this section,
for completeness, we repeat its derivation for a three-dimensional multiband
metal having in mind application to iron pnictides.

For the Hamiltonian (\ref{HamAF}) the collision integral due to scattering by
the spin fluctuations is related to the dynamic spin susceptibility $\chi_{j}
(\mathbf{q},\omega)$ as \cite{HlubRicePhysRevB94}
\begin{align}
&  \mathcal{I}_{s}(\mathbf{p})\!=\!\frac{g^{2}}{2}\!\int\frac{d\mathbf{p}
^{\prime} }{(2\pi)^{3}}\!\int\limits_{-\infty}^{\infty}\!d\omega\!\sum_{j}
\operatorname{Im}\chi_{j}(\mathbf{q},\omega) \delta\left(  \xi_{\bar
{s},\mathbf{p}^{\prime}}\!-\!\xi_{s,\mathbf{p}}\!+\!\omega\right) \nonumber\\
&  \times\left[  -\!f_{s,\mathbf{p}}(1\!-\!f_{\bar{s},\mathbf{p}^{\prime}}
)(n(\omega)\!+\!1)\!+\!f_{\bar{s},\mathbf{p}^{\prime}}(1\!-\!f_{s,\mathbf{p}}
)n(\omega)\right]  ,\label{CollIntegr}%
\end{align}
where $f_{s,\mathbf{p}}$ is the distribution function for the fermions in band
$s$, $\bar{s}\!=\!2(1)$ for $s\!=\!1(2)$, $\mathbf{q}=\mathbf{p}^{\prime
}\!-\!\mathbf{p}$, and $n(\omega)=\left[  \exp(\omega/T)-1\right]  ^{-1}$ is
the Bose-Einstein distribution function. For small deviations from
equilibrium, using standard presentation
\[
f_{s,\mathbf{p}}=f_{s,\mathbf{p}}^{0} -\frac{\partial f_{s,\mathbf{p}}^{0}
}{\partial\xi_{s,\mathbf{p}}} \Phi_{s,\mathbf{p}}=f_{s,\mathbf{p}}^{0}
+\frac{f_{s,\mathbf{p}}^{0}\left(  1-f_{s,\mathbf{p}} ^{0}\right)  }{T}
\Phi_{s,\mathbf{p}},
\]
we obtain
\begin{align}
\mathcal{I}_{s}(\mathbf{p})\!  &  =\!\frac{g^{2}}{2T}\!\int\frac
{d\mathbf{p}^{\prime}}{(2\pi)^{3}}\int\limits_{-\infty}^{\infty}\!d\omega\sum_{j}\operatorname{Im}
\chi_{j}(\mathbf{q},\omega)f_{s,\mathbf{p}}^{0}\left(  1\!-\!f_{\bar
{s},\mathbf{p}^{\prime}}^{0}\right) \nonumber\\
\times &  \left[  n(\omega)\!+\!1\right]  \left(  \Phi_{\bar{s},\mathbf{p}
^{\prime}}-\Phi_{s,\mathbf{p}}\right)  \delta\left(  \xi_{s,\mathbf{p}
}\!-\!\xi_{\bar{s},\mathbf{p}^{\prime} }\!-\!\omega\right)
,\label{CollIntLinear}%
\end{align}
where $f_{s,\mathbf{p}}^{0}=\left[  \exp(\xi_{s,\mathbf{p}}/T)+1\right]
^{-1}$ is the Fermi-Dirac distribution function.

The collision integral can be simplified using the standard transformation, $\int
d\mathbf{p}^{\prime}\rightarrow\int\frac{dS_{\bar{s}}^{\prime}} {|v_{\bar{s} }^{\prime}|}\int
d\xi_{\bar{s},\mathbf{p}^{\prime}}$, where $\int dS_{\bar{s}}^{\prime}\ldots$ means the integral over the
Fermi surface of $\bar{s}$ band and $v_{\bar{s}}^{\prime}$ is the Fermi velocity for this band.
Assuming that $\Phi_{\bar{s},\mathbf{p}^{\prime}}$ changes weakly on the scale $\xi_{\bar
	{s},\mathbf{p}^{\prime}}\sim T$, one can perform the energy integration independently, which allows
us to reduce $\mathcal{I}_{s}(\mathbf{p})$ to the following form
\begin{align*}
\mathcal{I}_{s}(\mathbf{p})  &  =-\frac{g^{2}}{2(2\pi)^{3}}\!\frac{\partial
f_{s,\mathbf{p}}^{0}}{\partial\xi_{s,\mathbf{p}}}\\
\times &  \int\frac{dS_{\bar{s} }^{\prime}}{|v_{\bar{s}}^{\prime}|}\left(
\Phi_{\bar{s},\mathbf{p}^{\prime} }\!-\!\Phi_{s,\mathbf{p}}\right)  \sum_{j}
K_{j}(\mathbf{q},\xi_{s,\mathbf{p}})
\end{align*}
with
\[
K_{j}(\mathbf{q},\xi)=\int d\xi^{\prime}\operatorname{Im}\chi_{j}
(\mathbf{q},\xi-\xi^{\prime} )\frac{\cosh(\frac{\beta\xi}{2})}{2\cosh
(\frac{\beta\xi^{\prime}} {2})\sinh\left[  \frac{\beta\left(  \xi
\!-\!\xi^{\prime}\right)  }{2}\right]  },
\]
where we used the following relations
\[
\left[  1\!-\!f^{0}(\xi^{\prime})\right]  \left[  n\left(  \xi\!-\!\xi
^{\prime}\right)  +1\right]  \!=\!\frac{\left[  1\!-\!f^{0}(\xi)\right]
\cosh(\frac{\beta\xi}{2})}{2\cosh(\frac{\beta\xi^{\prime}}{2})\sinh\left[
\frac{\beta\left(  \xi\!-\!\xi^{\prime}\right)  }{2}\right]  }%
\]
and $df^{0}/d\xi=-f^{0}(1-f^{0})/T$.

\begin{figure}[ptb]
\includegraphics[width=3.4in]{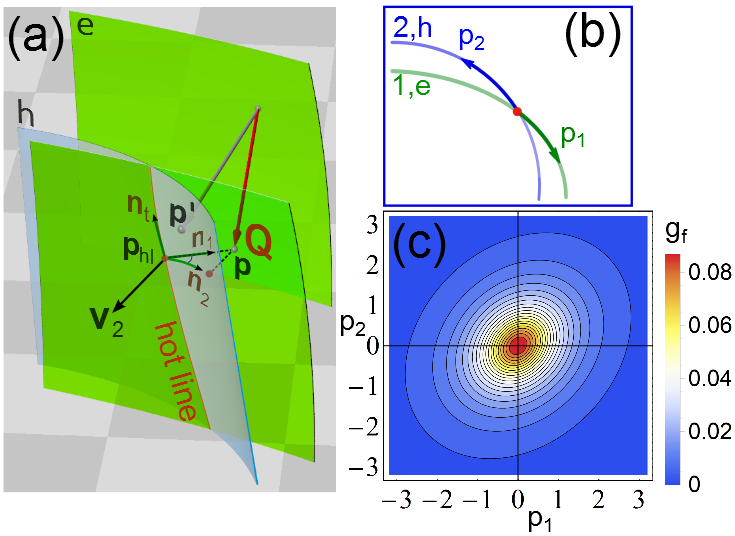}
\caption{(a) Fermi-surface geometry near the hot line. Projection of the electron
	Fermi-surface section displaced by the AF wave vector $\mathbf{Q}$ intersects the hole
	section along the hot line. For scattering event, the initial momentum $\mathbf{p}$ is
	located at the displaced electron Fermi surface and final momentum $\mathbf{p}^{\prime}$
	is located at the hole Fermi surface. The momentum $\mathbf{p}_{\mathrm{hl}}$ marks the
	location on the hot line closest to $\mathbf{p}$. (b) Cross sections of the hole and
	electron Fermi surfaces intersecting at the hot spot. The momenta $p_{s}$ with $s=1,2$
	measure distances from the hot spot along the corresponding Fermi surfaces. 
	(c) Representative contour plot of the transition rate
	$g_{\mathrm{f}}(p_{s},p_{\bar{s}})$, Eq.\ (\ref{g-ps-ps}). We assumed that all $\alpha_j$
	are identical and used the following parameters, $\eta_1=\eta_2$, $\eta_{12}=0.25\eta_1$,
	$(\pi/2)T\gamma=0.25\alpha$. $p_s$ and $g_f$ are measured in units of
	$\sqrt{\alpha/\eta_1}$ and $(3g^{2}T)/(16\pi \sqrt{\eta _{t}\alpha })$, respectively. } %
\label{Fig-ScatterFS}%
\end{figure}
We consider a quasiparticle at the Fermi level, $\xi
_{s,\mathbf{p}}=0$. In this case, with the shape of susceptibility given by
Eq.\ (\ref{GaussSuscept}), the energy integration reduces to calculation of
the reduced integral
\[
\zeta\left(  a\right)  =\int_{-\infty}^{\infty}du\frac{u}{a^{2}+u^{2}}
\frac{1}{\sinh u}.
\]
We can approximate this integral by the interpolation formula
\[
\zeta\left(  a\right)  \approx\frac{\pi}{a(1+2a/\pi)},
\]
which correctly reproduces its asymptotics. In this case $K_{j}(\mathbf{q},0)
$ takes the following form
\[
K_{j}(\mathbf{q},0)\approx\frac{\frac{\pi^{2}}{2}T^{2}\gamma}{\left(
\alpha_{j} +\sum_{i} \eta_{i}q_{i}^{2}\right)  (\frac{\pi}{2}T\gamma
+\alpha_{j}+\sum_{i}\eta_{i}q_{i}^{2})},
\]
and we obtain a useful intermediate result for the collision integral
\begin{align}
&  \mathcal{I}_{s}(\mathbf{p})\!\approx\!-\frac{g^{2}}{2(2\pi)^{3}}
\!\frac{\partial f_{s,\mathbf{p}}^{0}}{\partial\xi_{s,\mathbf{p}}}\nonumber\\
&  \times\sum_{j}\int\frac{dS_{\bar{s}}^{\prime}}{|v_{\bar{s}}^{\prime}|}
\frac{\frac{\pi^{2}}{2}T^{2}\gamma\left(  \Phi_{\bar{s},\mathbf{p}^{\prime}
}-\Phi_{s,\mathbf{p}}\right)  } {\left(  \alpha_{j}\!+\!\sum_{i}\eta_{i}
q_{i}^{2}\right)  (\frac{\pi}{2}T\gamma+\alpha_{j}\!+\!\sum_{i}\eta_{i}
q_{i}^{2})},
\end{align}
which contains two-dimensional integration over the Fermi surface. Further
simplification can be done observing that $\Phi_{\bar{s},\mathbf{p}^{\prime}}
$ strongly depends on the distance between $\mathbf{p}^{\prime}$ and the hot
line but varies smoothly along this line. Therefore one can perform
integration over the component $\mathbf{p}^{\prime}$ along the hot line
neglecting the dependence of $\Phi_{\bar{s},\mathbf{p}^{\prime}}$ on this
component \cite{RoschPhysRevB00}. This integration in general case, however,
is somewhat complicated by the anisotropy of the susceptibility characterized
by the parameters $\eta_{i}$. To proceed, we introduce the unit vector along
the hot line $\mathbf{n}_{t}$ and the unit vectors along the electron and hole
Fermi surfaces $\mathbf{n}_{s}$ with $s=1,2$ satisfying the conditions
$\sum_{i}\eta_{i}n_{t,i}n_{s,i}=0$, which replace the orthogonality conditions
in the isotropic case. Geometry around the hot line is illustrated in
Fig.\ \ref{Fig-ScatterFS}(a). We can now decompose the momenta $\mathbf{p}$
and $\mathbf{p}^{\prime}$ as $\mathbf{p} =\mathbf{p}_{\mathrm{hl} }+\mathbf{n}
_{s}p_{s}$, $\mathbf{p}^{\prime}=\mathbf{p}_{\mathrm{hl} }+\mathbf{n}_{t}
p_{t}+\mathbf{n}_{\bar{s}}p_{\bar{s}}$, where $\mathbf{p} _{\mathrm{hl}}$ is
the hot-line momentum closest to $\mathbf{p}$ and $p_{t}$ is the component of
$\mathbf{p}^{\prime}$ along the hot-line. The momentum components $p_{s}$
measure distance to the hot line, see Fig.\ \ref{Fig-ScatterFS}(b). With such
a decomposition, the sum $\sum_{i} \eta_{i}q_{i}^{2}$ takes the following
form
\[
\sum_{i}\eta_{i}q_{i}^{2}=\sum_{i}\eta_{i}\left(  p_{i}-p_{i}^{\prime}\right)
^{2}=\eta_{t}p_{t}^{2}+\eta_{s}p_{s}^{2}+\eta_{\bar{s}}p_{\bar{s}}^{2}
-2\eta_{s\bar{s}}p_{s}p_{\bar{s}}%
\]
with
\[
\eta_{t}=\sum_{i}\eta_{i}n_{t,i}^{2},\ \eta_{s}=\sum_{i}\eta_{i}n_{s,i}
^{2},\text{and }\eta_{s,\bar{s}}=\sum_{i}\eta_{i}n_{s,i}n_{\bar{s},i}.
\]
As the distribution function $\Phi_{\bar{s},\mathbf{p}^{\prime}}$ only weakly
depends on the parallel momentum $p_{t}$, we can neglect this dependence,
$\Phi_{\bar{s},\mathbf{p}^{\prime}}\rightarrow\Phi_{\bar{s}}(p_{\bar{s}})$,
and perform integration over $p_{t}$ in $\mathcal{I}_{s}(\mathbf{p})$ which
leads us to the following final presentation of the collision integral
\begin{equation}
\mathcal{I}_{s}(p_{s})\!\approx\!-\!\frac{\partial f_{s,\mathbf{p}}^{0}
}{\partial\xi_{s,\mathbf{p}}}\int\frac{dp_{\bar{s}}}{|v_{\bar{s}}|}
g_{\mathrm{f}}(p_{s},p_{\bar{s}})\left[  \Phi_{\bar{s}}(p_{\bar{s}})-\Phi_{s}
(p_{s})\right] \label{CollIntFinal}%
\end{equation}
with
\begin{align}
g_{\mathrm{f}}(p_{s},p_{\bar{s}})  &  =\frac{g^{2}T}{16\pi\sqrt{\eta_{t}}}
\sum_{j} \left(  \frac{1}{\sqrt{ \alpha_{j}\!+\!u(p_{s},p_{\bar{s}}) }}
\right. \nonumber\\
-  &  \left.  \frac{1}{\sqrt{ \frac{\pi} {2}T\gamma\!+\!\alpha_{j}
\!+\!u(p_{s},p_{\bar{s}}) }} \right)  ,\label{g-ps-ps}\\
u(p_{s},p_{\bar{s}})  &  =\eta_{s}p_{s}^{2}+\eta_{\bar{s}}p_{\bar{s}}^{2}
-2\eta_{s\bar{s}}p_{s}p_{\bar{s}}.\nonumber
\end{align}
The transition rate $g_{\mathrm{f}}(p_{s},p_{\bar{s}})$ increases as both the initial and final
momenta approach the hot line, $p_{s},p_{\bar{s}}\rightarrow 0$. Its shape is determined
by the parameters of dynamic susceptibility, $\alpha_j$, $\eta_i$, and $\gamma$, see Eq.\
(\ref{GaussSuscept}). The first term in parentheses in Eq.\ (\ref{g-ps-ps}) describes
elastic scattering by static ``snapshots'' of the fluctuating magnetization, while the
second term gives dynamic inelastic contribution. The typical behavior of
$g_{\mathrm{f}}(p_{s},p_{\bar{s}})$ is illustrated by the contour plot in Fig.\
\ref{Fig-ScatterFS}(c).
Equations (\ref{CollIntFinal}) and (\ref{g-ps-ps}) give the simplest accurate presentation for the collision
integral which can be used for precise numerical calculations of transport properties.

\begin{figure}[pt]
	\includegraphics[width=3.0in]{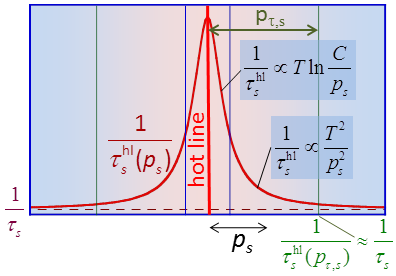}
	\caption{Behavior of the quasiparticle lifetime, Eq.\ (\ref{QPLifetime}), near the hot line.
	}
	\label{Fig-QPLifetime}%
\end{figure}
The intensity of scattering near the hot line can be characterized by the
quasiparticle lifetime $\tau_{s}^{\mathrm{hl}}(p_{s})$,
\begin{align}
\frac{1}{\tau_{s}^{\mathrm{hl}}}(p_{s})  &  =\int\frac{dp_{\bar{s}}}
{|v_{\bar{s}}|}g_{\mathrm{f}}(p_{s},p_{\bar{s}})\nonumber\\
&  =\frac{g^{2}T}{16\pi\sqrt{\eta_{t}\eta_{\bar{s}}}|v_{\bar{s}}|}\sum_{j}
\ln\left(  1+\frac{\frac{\pi}{2}T\gamma}{\alpha_{j}+\widetilde{\eta}_{s}
p_{s}^{2} }\right) \label{QPLifetime}%
\end{align}
with
\[
\widetilde{\eta}_{s}\equiv\frac{\eta_{s}\eta_{\bar{s}}-\eta_{s\bar{s}}^{2}
}{\eta_{\bar{s}}}.
\]
Behavior of the quasiparticle lifetime near the hot line is illustrated in Fig.\
\ref{Fig-QPLifetime}. One can see that there are two regimes of scattering, depending on
the temperature and proximity to the hot line. For $\gamma T\ll\widetilde{\eta
}_{s}p_{s}^{2}$ the scattering rate behaves as $1/\tau_{s}^{\mathrm{hl} }\propto
T^{2}/p_{s}^{2}$. For $\gamma T\gg\widetilde{\eta}_{s}p_{s}^{2}$ the scattering rate grows
logarithmically $1/\tau_{s}^{\mathrm{hl}}\propto T/\ln(C/|p_{s}|)$ and saturates at
$|p_{s}| =\sqrt{\alpha_{j}/\widetilde{\eta }_{s}}$. In the latter regime the frequency
dependence of the susceptibility is not essential meaning that this regime corresponds to
scattering on static ``frozen'' AF fluctuations. In general, the $\eta$-parameters depend
on local orientation of the hot line. For common particular case of hot line oriented
along the $z$ axis and no in-plane anisotropy, $\eta_{x}=\eta_{y}$, we have simple
relations $\eta_{t}=\eta_{z}$, $\eta_{s}=\eta_{\bar{s}}=\eta_{x}$,
$\eta_{s\bar{s}}=\eta_{x}\cos\alpha_{\mathrm{eh}}$, and $\widetilde{\eta}
_{s}=\eta_{x}\sin^{2}\alpha_{\mathrm{eh}}$, where $\alpha_{\mathrm{eh}}$ is the angle
between the electron and hole Fermi surfaces. In this case $u(p_{s},p_{\bar{s}})$ in Eq.\
(\ref{g-ps-ps}) is just proportional to the momentum change squared, 
$u(p_{s},p_{\bar{s}})=\eta_{x}(p_s\mathbf{n}_s-p_{\bar{s}}\mathbf{n}_{\bar{s}})^2$.

\section{Solution of Boltzmann equation using approximate transition rates.}
\label{Sec:SolBoltz}

To obtain conductivity in magnetic field, one has to solve the Boltzmann
kinetic equation for the distribution function \cite{ZimanBook,BlattBook}. We
assume that the electric field $\mathbf{E}$ is applied in the $xy$ plane and
the magnetic field $H$ applied along $z$ axis. Using simplified collision
integral, Eqs.\ (\ref{CollIntFinal}) and (\ref{g-ps-ps}), the two-band
Boltzmann equation takes the following approximate form
\begin{align}
-  &  eE_{\alpha}v_{s,\alpha}-\frac{e}{c}Hv_{s}\frac{d\Phi_{s,\alpha}}{dp_{s}
}\!=\!-\frac{\Phi_{s,\alpha}}{\tau_{s}}\nonumber\\
&  +\int\frac{dp_{s}}{|v_{s}|}g_{\mathrm{f}}(p_{s},p_{\bar{s}})\left[
\Phi_{\bar{s} ,\alpha}(p_{\bar{s}})-\Phi_{s,\alpha}(p_{s})\right]
,\label{Boltzmann}%
\end{align}
where $\tau_{s}$ are the background scattering times, which we assume to be
isotropic. We remind that in our notations $s=1$ corresponds to the electron
band, $v_{1}>0$, and $s=2$ corresponds to the hole band, $v_{2}<0$. Equation
(\ref{Boltzmann}) represents a system of coupled one-dimensional
integro-differential equations for the distribution functions depending on
distances from the hot spot $p_{s}$, see Fig.\ \ref{Fig-ScatterFS}(b). These
equations do not have exact analytical solution. Therefore, one can either
solve them numerically or rely on some approximations. The most common
approach is the relaxation-time approximation within which the return
scattering events described by the term $\int\frac{dp_{s}}{|v_{s} |}%
g(p_{s},p_{\bar{s}} )\Phi_{\bar{s},\alpha}(p_{\bar{s}})$ are completely
neglected. Even though in most cases this approximation gives physically
reasonable predictions, for strongly anisotropic scattering, it is not quantitatively accurate
\cite{KontaniRPP08,BreitkreizPhysRevB14}. Alternatively, the precise solution
of the kinetic equation can be obtained numerically. We propose a different
approximate scheme, which also allows for exact analytical solution and
preserves several realistic properties of the system which are lost in the
relaxation-time model.
We will replace the exact transition rates (\ref{g-ps-ps}) with the
approximate factorizable form
\begin{equation}
g_{\mathrm{f}}(p_{1},p_{2})\!=\!\gamma_{\mathrm{hs}}\psi_{1}(p_{1})\psi
_{2}(p_{2}),
\label{FactForm}
\end{equation}
where the functions $\psi_{s}(p_{s})$ are normalized as
\[
\int_{-\infty}^{\infty}\frac{dp_{s}}{|v_{s}|}\psi_{s}(p_{s})=1
\]
and their shapes are chosen to reproduce the hot-line relaxation time
(\ref{QPLifetime}),
\begin{equation}
\frac{1}{\tau_{s}^{\mathrm{hl}}(p_{s})}=\gamma_{\mathrm{hs}}\psi_{s}%
(p_{s}).\label{taupsi}%
\end{equation}
Therefore, the total amplitude characterizing the strength of hot-spot
scattering is given by
\[
\gamma_{\mathrm{hs}}=\int\frac{dp_{s}}{|v_{s}|}\frac{1}{\tau_{s}^{\mathrm{hl}
}(p_{s})}=\int\frac{dp_{1}}{v_{1}}\int\frac{dp_{2}}{|v_{2}|}g_{\mathrm{f}
}(p_{1},p_{2}).
\]
The relative strength of the hot spot with respect to background scattering
can be conveniently characterized by the reduced parameter $\gamma
_{\mathrm{hs}}|v_{s}|\tau_{s}/p_{F,s}$. We will assume that this parameter is
small.

Introducing notations
\begin{subequations}
\begin{align}
&  \Phi_{s,\alpha}(p_{s})\!=\!eE_{\alpha}\Lambda_{s,\alpha}(p_{s}),
\label{NotatLambda}\\
&  \bar{\Lambda}_{s,\alpha}\!=\!\int\frac{dp_{s}}{|v_{s}|}\psi_{s}
(p_{s})\Lambda_{s,\alpha}(p_{s}),\label{DefLambdaBar}%
\end{align}
we obtain from Eq.\ (\ref{Boltzmann}) the following equation for the ``vector mean-free path''\cite{Taylor200PRSA63} 
$\Lambda_{s,\alpha}(p_{s})$ 
\end{subequations}
\begin{equation}
\left[  \frac{1}{\tau_{s}}\!+\!\gamma_{\mathrm{hs}}\psi_{s}(p_{s})\right]
\Lambda_{s,\alpha}\!-\!\frac{e}{c}Hv_{s}\frac{d\Lambda_{s,\alpha}}{dp_{s}
}\!=\!v_{s,\alpha}\!+\!\gamma_{\mathrm{hs}}\psi_{s}(p_{s})\bar{\Lambda}
_{\bar{s},\alpha}.\label{EqLambda}%
\end{equation}
The conductivity tensor is related to $\Lambda_{s,\alpha}(p_{s})$ as
\begin{subequations}
\begin{align}
\sigma_{\alpha\beta}  &  =\frac{2e^{2}}{(2\pi)^{3}}\int dp_{z}S_{\alpha\beta
},\label{sigma-S}\\
S_{\alpha\beta}  &  \approx\sum_{s}\int\frac{dp_{s}}{|v_{s}|}v_{s,\alpha
}\Lambda_{s,\beta}.\label{CondLambd}%
\end{align}
The physical parameters $v_{s}$ and $\gamma_{\mathrm{hs}}$ entering
Eq.\ (\ref{EqLambda})
depend on $z$ axis momentum $p_{z}$ as an external parameter. Therefore this
equation deals with fixed-$p_{z}$ cross section of the Fermi surface which
intersects hot lines at the hot spots. The total conductivity is obtained by
integration over $p_{z}$. In the following, we will skip an implicit dependence
on $p_{z}$ in all parameters. The hot-line contribution to the conductivity,
$\sigma_{\alpha\beta}^{\mathrm{hl}}$, is given by
\end{subequations}
\begin{equation}
\sigma_{\alpha\beta}^{\mathrm{hl}} =\frac{2e^{2}}{(2\pi)^{3}}\int dp_{z}%
\sum_{\mathrm{hs}} S_{\alpha\beta}^{\mathrm{hs}},\label{sigmahl}%
\end{equation}
where $S_{\alpha\beta}^{\mathrm{hs}}$ is the contribution to $S_{\alpha\beta}
$ from one hot spot and the sum is taken over all hot spots in the given
$p_{z} $ cross section. In the following sections we solve
Eq.\ (\ref{EqLambda}) and compute components of conductivity.

\section{Conductivity in zero magnetic field}

\label{Sec:CondZeroH}

For completeness, we consider first the conductivity at zero magnetic field,
see also Ref.\ \onlinecite{StojkovicPinesPhysRevB97}. Rewriting
Eq.\ (\ref{EqLambda}) as
\[
\Lambda_{s,\alpha}\!=\!\frac{v_{s,\alpha}}{\frac{1}{\tau_{s}}+\gamma
_{\mathrm{hs}}\psi_{s}(p_{s})}+\frac{\gamma_{\mathrm{hs}}\psi_{s}(p_{s}
)}{\frac{1}{\tau_{s}}+\gamma_{\mathrm{hs}}\psi_{s}(p_{s})}\bar{\Lambda}
_{\bar{s},\alpha},
\]
we obtain $2\times2$ linear system for $\bar{\Lambda}_{s,\alpha}$ defined by
Eq.~(\ref{DefLambdaBar})
\begin{subequations}
\label{EqLambdaBar}%
\begin{align}
\bar{\Lambda}_{1,\alpha}-\left(  1-r_{1}\right)  \bar{\Lambda}_{2,\alpha}\!
&  =r_{1}\tau_{1}v_{1,\alpha}^{\mathrm{hs}},\\
-\left(  1-r_{2}\right)  \bar{\Lambda}_{1,\alpha}+\bar{\Lambda}_{2,\alpha}\!
&  =r_{2}\tau_{2}v_{2,\alpha}^{\mathrm{hs}},
\end{align}
where $v_{s,\alpha}^{\mathrm{hs}}$ are Fermi-velocity components at the hot
spot and the dimensionless parameters $r_{s}$ are defined as
\end{subequations}
\begin{equation}
r_{s}\equiv\int\!\frac{\psi_{s}(p_{s})dp_{s}/|v_{s}|}{1+\tau_{s}
\gamma_{\mathrm{hs}}\psi_{s}(p_{s})}=\frac{1}{\gamma_{\mathrm{hs}}}\int%
\!\frac{dp_{s}/|v_{s}|}{\tau_{s}+\tau_{s}^{\mathrm{hl}}(p_{s})}.\label{rs}%
\end{equation}
The solution of these equations is
\begin{equation}
\bar{\Lambda}_{s,\alpha}=\frac{r_{s}\tau_{s}v_{s,\alpha}^{\mathrm{hs}}
+r_{\bar{s}}\left( 1-r_{s}\right) \tau_{\bar{s}}v_{\bar{s},\alpha
}^{\mathrm{hs}}} {r_{1}+r_{2}-r_{1}r_{2}}.\label{LambBar}%
\end{equation}
In the case of narrow hot spot,\ $r_{s} \ll1$, we obtain the following
approximate result for the vector mean-free path
\begin{align}
&  \Lambda_{s,\alpha}(p_{s})\!\approx\frac{r_{1}v_{1,\alpha}^{\mathrm{hs}}
\tau_{1}+r_{2}v_{2,\alpha}^{\mathrm{hs}}\tau_{2}}{r_{1}+r_{2}}\nonumber\\
&  +\frac{1}{1\!+\!\tau_{s}/\tau_{s}^{\mathrm{hl}}(p_{s})}
\left( v_{s,\alpha}\tau_{s}\!-\frac{r_{1}v_{1,\alpha}^{\mathrm{hs}}\tau_{1}
	\!+\!r_{2}v_{2,\alpha}^{\mathrm{hs}}\tau_{2}}{r_{1}+r_{2}}\right).
\label{LambdZeroH}%
\end{align}
Here the first term approximately gives the vector mean-free path in the
hot-spot region. Due to strong equilibration, it is identical for two
bands.\footnote{Note that, in contrast to the relaxation-time approximation,
the distribution function does not vanish in the hot-spot region.} This result
can also be rewritten approximately as \footnote{As the second term vanishes
away from the hot spot, in its derivation we neglected $p_{s}$ dependence of
$v_{s,\alpha}$ and replaced $v_{s,\alpha}\rightarrow v_{s,\alpha}
^{\mathrm{hs}}$.}
\begin{equation}
\Lambda_{s,\alpha}(p_{s})\!\approx v_{s,\alpha}\tau_{s}\!+\!\frac{\tau_{s}
}{\tau_{s}\!+\!\tau_{s}^{\mathrm{hl}}(p_{s})}\frac{\left(  v_{\bar{s},\alpha
}^{\mathrm{hs}}\tau_{\bar{s}}\!-v_{s,\alpha}^{\mathrm{hs}}\tau_{s}\right)
r_{\bar{s}}}{r_{1}+r_{2}}.\label{LambdZeroH1}%
\end{equation}
We can see that far away from the hot spot where $\tau_{s}^{\mathrm{hl}
}(p_{s})\gg\tau_{s}$, the conventional result $\Lambda_{s,\alpha}
(p_{s})\approx v_{s,\alpha}\tau_{s}$ is restored. The transition to this
asymptotic takes place at the typical momentum $p_{\tau,s}$ where the hot-spot
scattering rate drops down to the background, $\tau_{s}^{\mathrm{hl}}
(p_{\tau,s})=\tau_{s}$, see Fig.\ \ref{Fig-QPLifetime}. Note that this typical momentum is mostly determined
by the tail region in the scattering rate $1/\tau_{s}^{\mathrm{hl}}(p_{s})$
and changes only weakly when the temperature approaches the transition point.

Substituting result from Eq.\ (\ref{LambdZeroH1}) into Eq.\ (\ref{CondLambd}),
we obtain
\begin{subequations}
\begin{equation}
S_{\alpha\alpha}=S_{\alpha\alpha}^{(0)}+S_{\alpha\alpha}^{\mathrm{hs}%
},\label{SzeroHt}%
\end{equation}
with the background and hot-spot contributions given by
\begin{align}
S_{\alpha\alpha}^{(0)} & =\sum_{s}\int v_{s,\alpha}^{2}\tau_{s}\frac{dp_{s}%
}{|v_{s}|},\label{S0zeroH}\\
S_{\alpha\alpha}^{\mathrm{hs}} & =-\gamma_{\mathrm{hs}}\frac{\left(
v_{2,\alpha}^{\mathrm{hs}}\tau_{2}-v_{1,\alpha}^{\mathrm{hs}}\tau_{1}\right)
^{2}}{1/r_{1}+1/r_{2}}.\label{ShszeroH}%
\end{align}
Remind that $S_{\alpha\alpha}$ directly determines the conductivity
$\sigma_{\alpha\alpha}$ by Eq.\ (\ref{sigma-S}). Alternatively, the hot-spot
contribution can be expressed via the relaxation rates $\tau_{s}^{\mathrm{hl}%
}(p_{s})$,
\end{subequations}
\[
S_{\alpha\alpha}^{\mathrm{hs}}=-\left(  v_{2,\alpha}^{\mathrm{hs}}\tau
_{2}-v_{1,\alpha}^{\mathrm{hs}}\tau_{1}\right)  ^{2}\left\{  \sum_{s}\left[
\int\!\frac{dp_{s}/|v_{s}|}{\tau_{s}+\tau_{s}^{\mathrm{hl}}(p_{s})}\right]
^{-1}\right\}  ^{-1}%
\]
and can be estimated as
\[
S_{\alpha\alpha}^{\mathrm{hs}}\approx-\frac{\left(  v_{2,\alpha}^{\mathrm{hs}
}\tau_{2}-v_{1,\alpha}^{\mathrm{hs}}\tau_{1}\right)  ^{2}}{\frac{\tau
_{1}|v_{1}|}{p_{\tau,1}}+\frac{\tau_{2}|v_{2}|}{p_{\tau,2}}}.
\]
Typically, the cold regions dominate in transport and the hot spots give only
small corrections \cite{StojkovicPinesPhysRevB97}. Moreover, these corrections
are not singular at the transition point. As the carriers within the range
$\sim p_{\tau}$ from the hot line are almost eliminated from transport, the
relative reduction of conductivity is of the order of $\sigma_{\alpha\alpha
}^{\mathrm{hs}} /\sigma_{\alpha\alpha}^{(0)}\approx p_{\tau}/p_{F}$. Note,
however, that, contrary to the relaxation-time approximation, in the case
$r_{1}v_{1,\alpha}^{\mathrm{hs}}\tau_{1} +r_{2}v_{2,\alpha}^{\mathrm{hs}}
\tau_{2}\neq0$, the distribution functions do not vanish in the hot-spot
regions and therefore these regions actually give finite contributions to the
current and conductivity.

\section{Conductivity in magnetic field}
\label{Sec:CondMagnField}

In the magnetic field the formal solution of Eq.\ (\ref{EqLambda}) can be
written as
\begin{subequations}
\begin{align}
&  \Lambda_{s,\alpha}(p_{s})\!=\!\int\limits_{p_{s}}^{\delta_{s}\infty}
\!\frac{dp_{s}^{\backprime}}{v_{s}^{\backprime}}\frac{v_{s,\alpha}
^{\backprime}\!+\!\gamma_{\mathrm{hs}}\psi_{s}(p_{s}^{\backprime})\bar
{\Lambda}_{\bar{s},\alpha}}{\frac{e}{c}H}\mathcal{L}_{H,s}(p_{s}^{\backprime
},p_{s}),\label{Lam-ps-H}\\
&  \mathcal{L}_{H,s}(p_{s}^{\backprime},p_{s})\equiv\exp\left[  -\int_{p_{s}
}^{p_{s}^{\backprime}}\frac{\frac{1}{\tau_{s}}+\gamma_{\mathrm{hs}}\psi
_{s}(\tilde{p}_{s})}{\frac{e}{c}H}\frac{d\tilde{p}_{s}}{\tilde{v}_{s}}\right]
\label{LHsDef}%
\end{align}
with $\delta_{s}\equiv\mathrm{sgn}(ev_{s})\equiv-\mathrm{sgn}(v_{s})$. This
presentation is similar to so-called Shockley \textquotedblleft tube
integral\textquotedblright\cite{ShockleyPhysRev50}, see also
Ref.\ \onlinecite{AbdelJawadNPhys06} for the recent use of this approach. The
exponent in Eq.\ (\ref{LHsDef}) is the probability of reaching point
$p_{s}^{\backprime}$ from point $p_{s}$ without scattering during orbital
motion of quasiparticle in the magnetic field. The term with $\bar{\Lambda
}_{\bar{s},\alpha}$ in Eq.\ (\ref{Lam-ps-H}) describes the contribution from
the return-scattering events. Without this term Eq.\ (\ref{Lam-ps-H}) would
give the relaxation-time-approximation result. Using the identity
\end{subequations}
\[
\int_{p_{s}}^{\delta_{s}\infty}\!\frac{dp_{s}^{\backprime}}{v_{s}^{\backprime
}}\frac{\frac{1}{\tau_{s}}+\gamma_{\mathrm{hs}}\psi_{s}(p_{s}^{\backprime}
)}{\frac{e}{c}H}\mathcal{L}_{H,s}(p_{s}^{\backprime},p_{s})=1,
\]
we can also transform this presentation to the following form
\begin{equation}
\Lambda_{s,\alpha}(p_{s})\!=\!\bar{\Lambda}_{\bar{s},\alpha}+\int%
\limits_{p_{s}}^{\delta_{s}\infty}\!dp_{s}^{\backprime}\frac{\tau
_{s}v_{s,\alpha}^{\backprime}-\bar{\Lambda}_{\bar{s},\alpha}}{\frac{e}
{c}Hv_{s}^{\backprime}\tau_{s}}\mathcal{L}_{H,s}(p_{s}^{\backprime}
,p_{s}).\label{Lam-ps-H1}%
\end{equation}
From this result, we derive the linear system for the parameters $\bar
{\Lambda}_{s,\alpha}$,
\begin{equation}
\bar{\Lambda}_{s,\alpha}\!-\left(  1-R_{s}\right)  \bar{\Lambda}_{\bar
{s},\alpha}=\!\mathcal{V}_{s,\alpha}\tau_{s},\label{LamBarHEq}%
\end{equation}
where the parameters $R_{s}$ and $\mathcal{V}_{s,\alpha}$ are defined by the
double integrals,
\begin{subequations}
\begin{equation}
R_{s}=\!\frac{1}{\tau_{s}\gamma_{\mathrm{hs}}}\int\limits_{-\infty}^{\infty
}\frac{dp_{s}^{\backprime}}{|v_{s}^{\backprime}|}\!\int\limits_{-\delta
_{s}\infty}^{p_{s}^{\backprime}}\!\frac{dp_{s}}{\frac{e}{c}H\tau_{s}v_{s}
}\mathcal{M}_{H,s}(p_{s}^{\backprime},p_{s})\label{RsDef}
\end{equation}
and
\begin{equation}
\mathcal{V}_{s,\alpha}\!=\!\frac{1}{\tau_{s}\gamma_{\mathrm{hs}}}
\!\int\limits_{-\infty}^{\infty}\!v_{s,\alpha}^{\backprime}\frac
{dp_{s}^{\backprime}}{|v_{s}^{\backprime}|}\!\int\limits_{-\delta_{s}\infty
}^{p_{s}^{\backprime}}\!\frac{dp_{s}}{\frac{e}{c}H\tau_{s}v_{s}}
\mathcal{M}_{H,s}(p_{s}^{\backprime},p_{s})\label{Vsa}
\end{equation}
with
\begin{align}
\mathcal{M}_{H,s}& (p_{s}^{\backprime},p_{s})   \equiv\exp\left(
\!-\!\int_{p_{s}}^{p_{s}^{\backprime}}\frac{d\tilde{p}_{s}}{\frac{e}{c}
H\tau_{s}\tilde{v}_{s}}\right) \nonumber\\
\times &  \left[  1-\exp\left(  -\frac{\gamma_{\mathrm{hs}}}{\frac{e}{c}H}
\int_{p_{s}}^{p_{s}^{\backprime}}\!\psi_{s}(\tilde{p}_{s})\frac{d\tilde{p}_{s}
}{\tilde{v}_{s}}\right)  \right]. \label{MHsDef}
\end{align}
The solution of Eq.\ (\ref{LamBarHEq}) is
\end{subequations}
\begin{equation}
\bar{\Lambda}_{s,\alpha}=\frac{\!\mathcal{V}_{s,\alpha}\tau_{s}+\mathcal{V}
_{\bar{s},\alpha}\tau_{\bar{s}}\left(  1-R_{s}\right)  }{R_{1}+R_{2}
-R_{1}R_{2}}.\label{LamBarH}%
\end{equation}
This result determines the vector mean-free path by Eq.\ (\ref{Lam-ps-H1}),
which, in turn, determines the conductivity components by Eqs.\ (\ref{sigma-S})
and (\ref{CondLambd}). In the following sections we proceed with the derivation of
the longitudinal and Hall conductivities.

\subsection{Longitudinal conductivity}

\label{Sec:LongCond}

For calculation of the longitudinal conductivity, in the integral for
$\mathcal{V}_{s,\alpha}$, Eq.\ (\ref{Vsa}), one can replace $v_{s,\alpha
}^{\backprime}$ by its value at the hot line, $v_{s,\alpha}^{\mathrm{hs}}$,
giving $\mathcal{V}_{s,\alpha}\approx v_{s,\alpha}^{\mathrm{hs}}R_{s}$. To
proceed further, we need to obtain a tractable expression for the
field-dependent parameter $R_{s}(H)$, Eqs.\ (\ref{RsDef}) and (\ref{MHsDef}). The essential
magnetic field scale in this dependence, $B_{w,s}$, is set by the typical
width of the hot spot scattering rate $w_{s}\approx\sqrt{\alpha_{j}%
/\widetilde{\eta}_{s}}$ (the width of the functions $\psi_{s}(p_{s})$) as
\begin{equation}
B_{w,s}=\frac{c}{|e|}\sqrt{\frac{\pi\gamma_{\mathrm{hs}}w_{s}}{|v_{s}%
^{\mathrm{hs}}|\tau_{s}}}.\label{Bw}%
\end{equation}
While behavior at very small magnetic fields, $H\ll B_{w,s}$, is sensitive to
exact shape of $\psi_{s}(p_{s})$, at higher fields the internal structure of
$\psi_{s}(p_{s})$ is not important and it can be treated as $\delta
$-functions, $\psi_{s}(p_{s})\rightarrow|v_{s}|\delta(p_{s})$. This allows us
to derive relatively simple analytical results for this magnetic field regime.
In this case, we obtain that the parameters $R_{s}$, Eq.\ (\ref{RsDef}), are
identical for both bands and given by
\begin{equation}
R_{s}\approx\frac{H}{B_{\gamma}}\left[  1\!-\!\exp\left(  -\frac{B_{\gamma}%
}{H}\right)  \right]  \label{RsAsymp}%
\end{equation}
with the field scale
\begin{equation}
B_{\gamma}=\frac{c}{|e|}\gamma_{\mathrm{hs}}\label{Bgam}%
\end{equation}
set by the total scattering amplitude. The exponential factor in this result
represents probability for a quasiparticle to pass through the hot spot
without scattering. For $H\ll B_{\gamma}$ this probability is negligibly
small. Substituting result (\ref{RsAsymp}) into Eqs.\ (\ref{LamBarH}), we
obtain
\[
\bar{\Lambda}_{s,\alpha}\!\approx\!\frac{v_{s,\alpha}^{\mathrm{hs}}\tau
_{s}+v_{\bar{s},\alpha}^{\mathrm{hs}}\tau_{\bar{s}}\left\{  1-\frac
{H}{B_{\gamma}}\left[  1\!-\!\exp\left(  -\frac{B_{\gamma}}{H}\right)
\right]  \right\}  }{\!2\!-\!\frac{H}{B_{\gamma}}\left[  1\!-\!\exp\left(
-\frac{B_{\gamma}}{H}\right)  \right]  }%
\]
and from Eq.\ (\ref{Lam-ps-H1}) the vector mean-free path
\begin{align}
&  \Lambda_{s,\alpha}(p_{s})\!\approx\!v_{s,\alpha}\tau_{s}-\left(
v_{s,\alpha}^{\mathrm{hs}}\tau_{s}-v_{\bar{s},\alpha}^{\mathrm{hs}}\tau
_{\bar{s}}\right)  \theta(-\delta_{s}p_{s})\nonumber\\
&  \times\exp\left(  \frac{p_{s}}{\frac{e}{c}Hv_{s}^{\mathrm{hs}}\tau_{s}%
}\right)  \frac{1\!-\!\exp\left(  -\frac{B_{\gamma}}{H}\right)  }%
{2\!-\!\frac{H}{B_{\gamma}}\left[  1\!-\!\exp\left(  -\frac{B_{\gamma}}%
{H}\right)  \right]  },\label{LamLinear}%
\end{align}
where $\theta(x)$ is the step function. We can see that the hot spot affects
the distribution function only on one side, in the range $\Delta_{H}%
p_{s}=\frac{e}{c}H|v_{s}^{\mathrm{hs}}|\tau_{s}$, meaning that the affected
area of the Fermi surface grows proportionally to the magnetic field. This
result also means that the hot spots influence conductivity independently
until $\Delta_{H}p_{s}<p_{F,s}$. Substituting derived $\Lambda_{s,\alpha
}(p_{s})$ into Eq.\ (\ref{CondLambd}), we obtain the magnetic-field dependent
part of $S_{\alpha\alpha}^{\mathrm{hs}}(H)$, $\delta S_{\alpha\alpha
}^{\mathrm{hs}}(H)\equiv S_{\alpha\alpha}^{\mathrm{hs}}(H)-\!S_{\alpha\alpha
}^{\mathrm{hs}}(0)$,
\begin{equation}
\delta S_{\alpha\alpha}^{\mathrm{hs}}(H)\!\approx\!-\left(  v_{1,\alpha
}^{\mathrm{hs}}\tau_{1}\!-\!v_{2,\alpha}^{\mathrm{hs}}\tau_{2}\right)
^{2}\!\frac{\frac{e}{c}H\left[  1\!-\!\exp\left(  -\frac{B_{\gamma}}%
{H}\right)  \right]  }{2\!-\!\frac{H}{B_{\gamma}}\left[  1\!-\!\exp\left(
-\frac{B_{\gamma}}{H}\right)  \right]  },\label{Saa-HighH}%
\end{equation}
meaning that the reduction of conductivity due hot-line scattering increases
\emph{linearly} with the magnetic field within $B_{w,s}<H<B_{\gamma}$,
\begin{equation}
\delta S_{\alpha\alpha}^{\mathrm{hs}}\approx-\frac{1}{2}\left(  v_{1,\alpha
}^{\mathrm{hs}}\tau_{1}\!-\!v_{2,\alpha}^{\mathrm{hs}}\tau_{2}\right)
^{2}\frac{e}{c}H.\label{Sxx-hs-lin}%
\end{equation}
In this linear regime the penetration of a carrier through the hot spot
without scattering is negligible and the conductivity is not sensitive to the
hot-spot parameters at all. At higher field, $H>B_{\gamma}$, the hot-spot
contribution saturates at a finite value,
\begin{equation}
\delta S_{\alpha\alpha}^{\mathrm{hs}}\!\approx\!-\gamma_{\mathrm{hs}}\left(
v_{1,\alpha}^{\mathrm{hs}}\tau_{1}\!-\!v_{2,\alpha}^{\mathrm{hs}}\tau
_{2}\right)  ^{2}.\label{Sxx-hs-sat}%
\end{equation}
This result is valid assuming that the hot spots still act independently at
$H\sim B_{\gamma}$, which is correct if $B_{\gamma}<(c/e)p_{F,s}/|v_{s}%
|\tau_{s}$ corresponding to the condition for the hot-spot strength
$\gamma_{\mathrm{hs}}<p_{F,s}/|v_{s}|\tau_{s}$.

\begin{figure}[ptb]
\includegraphics[width=3.3in]{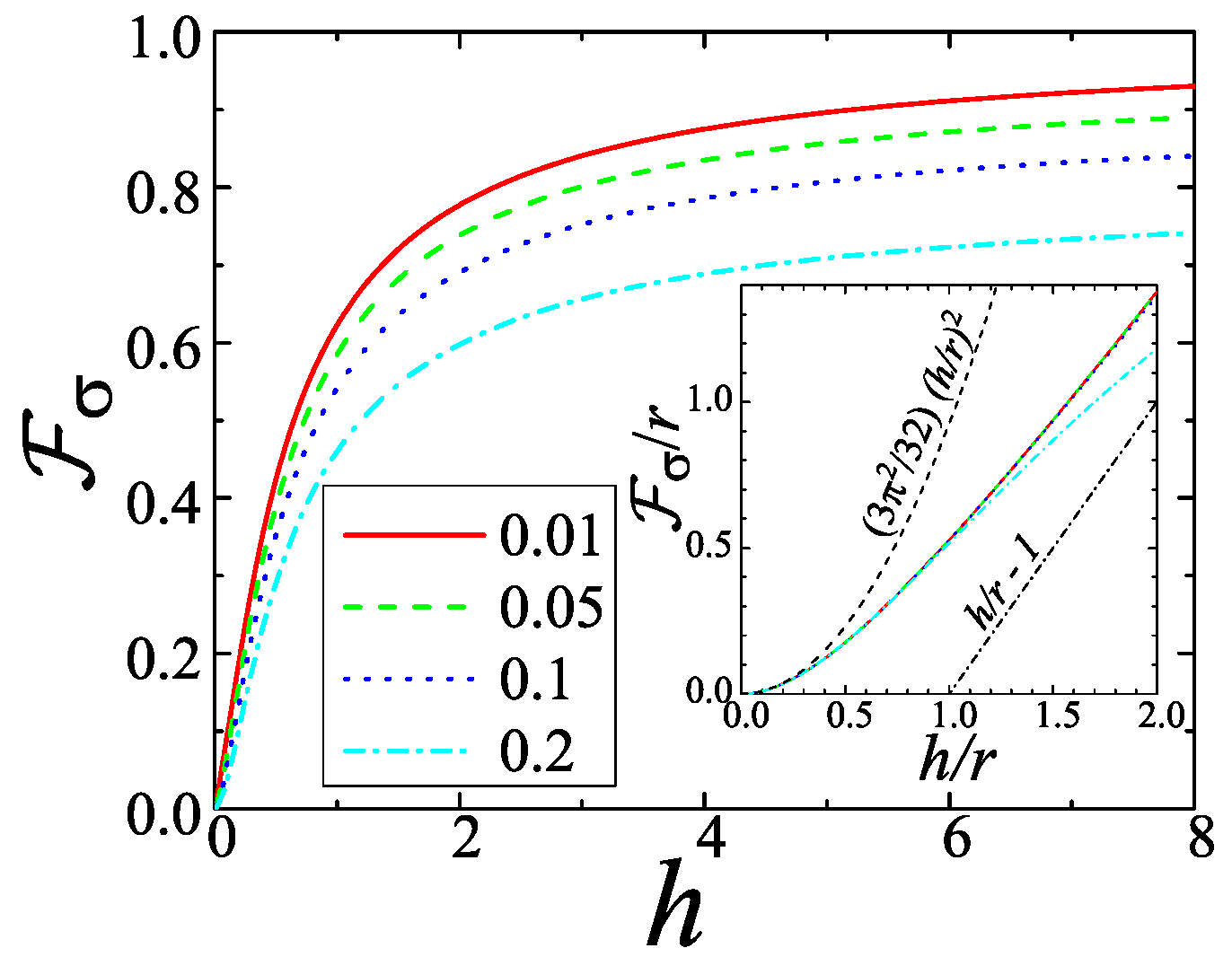}
\caption{The dependence of the function $\mathcal{F}_{\sigma}(h,r)$ defined by
Eqs.\ (\ref{Fs-h-r}) and (\ref{Fs-Present}) on the reduced field $h$ for
different values of $r$ specified in the legend. The inset shows the scaling
plot of $\mathcal{F}_{\sigma}/r$ vs $h/r $ describing the crossover between
the quadratic and linear regimes at small fields. The dashed and dot-dashed lines show
the low-field quadratic and linear asymptotics.}%
\label{Fig-Fs-h}%
\end{figure}

For quantitative description of the behavior in the full field range including
$B\sim B_{w,s}$, we assume a simple Lorentzian shape of $\psi_{s}(p_{s})$
valid for $T<\alpha_{j}/\gamma$, see Eq.\ (\ref{QPLifetime}), and close
$\alpha_{j}$ for all $j$,
\begin{equation}
\psi_{s}(p_{s})=\frac{|v_{s}^{\mathrm{hs}}|w_{s}/\pi}{p_{s}^{2}+w_{s}^{2}%
}.\label{LorentzShape}%
\end{equation}
Comparing with microscopic result, Eq.\ (\ref{QPLifetime}), we can express the
strength and width of the hot spot via the microscopic parameters as
\begin{equation}
\gamma_{\mathrm{hs}} =\frac{3\pi g^{2}\gamma T^{2}}{32|v_{\bar{s}}%
||v_{s}|\sqrt{\eta_{t}\left(  \eta_{s}\eta_{\bar{s}}-\eta_{s\bar{s}}%
^{2}\right)  \alpha_{x}}},\ w_{s}=\sqrt{\frac{\alpha_{x}}{\tilde{\eta}_{s}}}.
\end{equation}
In this case, for $w_{s}\ll\gamma_{\mathrm{hs}}\tau_{s}|v_{s}^{\mathrm{hs}
}|/\pi$, the parameter $r_{s}$, Eq.\ (\ref{rs}), can be evaluated as
\begin{equation}
r_{s}\approx\sqrt{\frac{\pi w_{s}}{\gamma_{\mathrm{hs}}|v_{s}^{\mathrm{hs}
}|\tau_{s}}}.\label{rsLorentz}%
\end{equation}
Note that this parameter also determines the ratio of the typical fields
$B_{w,s}$ and $B_{\gamma}$, $B_{w,s}=r_{s}B_{\gamma}$. The typical momentum
scale $p_{\tau,s}$ defined in the previous section by the condition $\tau
_{s}^{\mathrm{hl}}(p_{\tau,s})=\tau_{s}$ becomes $p_{\tau,s}=\sqrt
{\gamma_{\mathrm{hs}}|v_{s}^{\mathrm{hs}}|\tau_{s}w_{s}/\pi}=w_{s}/r_{s}$.
With such $\psi_{s}(p_{s})$ the function $\mathcal{M} _{H,s}(p_{s}%
^{\backprime},p_{s})$, Eq.\ (\ref{MHsDef}), can be evaluated analytically as
\begin{align}
&  \mathcal{M}_{H,s}(p_{s}^{\backprime},p_{s}) \approx\exp\left(
\!-\!\frac{p_{s}^{\backprime}\!-\!p_{s}}{\frac{e}{c}Hv_{s}\tau_{s}}\right)
\nonumber\\
&  \times\left\{  1\!-\!\exp\left[  -\frac{\gamma_{\mathrm{hs}}}{\frac{e}%
{c}\pi H}\left(  \arctan\frac{p_{s}^{\backprime}}{w_{s}}\!-\!\arctan
\frac{p_{s} }{w_{s}}\right)  \right]  \right\} .\label{MHLor}%
\end{align}
This allows us to transform the parameters $R_{s}$, Eq.\ (\ref{RsDef}), to the
following form
\begin{equation}
R_{s}=r_{s}+\mathcal{F}_{\sigma}(h,r_{s}),\label{Rs}%
\end{equation}
where $h=H/B_{\gamma}$ is the reduced magnetic field. The dimensionless
function $\mathcal{F}_{\sigma}(h,r_{s})$
is defined by the following double integral
\begin{align}
\mathcal{F}_{\sigma}(h,r)  &  \!=\!\frac{r^{2}}{\pi}\int_{-\infty}^{\infty
}\!du\int_{0}^{\infty}\!dz\exp\left(  -z\right)  \left\{  \exp\left[
-\frac{z/r^{2}}{1+u^{2}}\right]  \right. \nonumber\\
\!-\!  &  \left.  \exp\left[  -\frac{\arctan\left(  u+\frac{\pi hz}{r^{2}
}\right)  \!-\!\arctan u}{\pi h}\right]  \right\} \label{Fs-h-r}%
\end{align}
and has the following asymptotics
\[
\mathcal{F}_{\sigma}(h,r)\approx%
\genfrac{\{}{.}{0pt}{}{\frac{3\pi^{2}}{32}\frac{h^{2}}{r}\left(  1-\frac
{15\pi^{2}}{64}\frac{h^{2}}{r^{2}}\right)  ,\text{ for }h\ll r}{h\left[
1-\exp\left(  -\frac{1}{h}\right)  \right]  +r,\text{ for }h\gg r}%
.
\]
For the most typical case $r\ll1$ this function can be transformed to the form
with a single integration, see Appendix \ref{App-Fs-small-rApp},
\begin{align}
\mathcal{F}_{\sigma}(h,r)  &  =\frac{r}{\pi}\left[  1\!-\!\exp\left(
-\frac{1}{h}\right)  \right] \nonumber\\
&  \times\int_{0}^{\infty}dx\exp\left(  -x\right)  G\left(  \frac{\pi hx}
{4r}\right)  \!-\!r,\label{Fs-Present}\\
G(a)  &  =4\sqrt{1\!+\!a^{2}}E\left(  \frac{a^{2}}{1\!+\!a^{2}}\right)
-\frac{2}{\sqrt{1\!+\!a^{2}}}K\left(  \frac{a^{2}}{1\!+\!a^{2}}\right)
,\nonumber
\end{align}
where $E(m)$ and $K(m)$ are the full elliptic integrals. Figure \ref{Fig-Fs-h}
shows dependence of $\mathcal{F}_{\sigma}(h,r)$ on the reduced field $h$ for
different values of $r$. The inset shows crossover between the quadratic and
linear regimes at $h\sim r$.  %
\begin{figure}[ptb]
\includegraphics[width=3.3in]{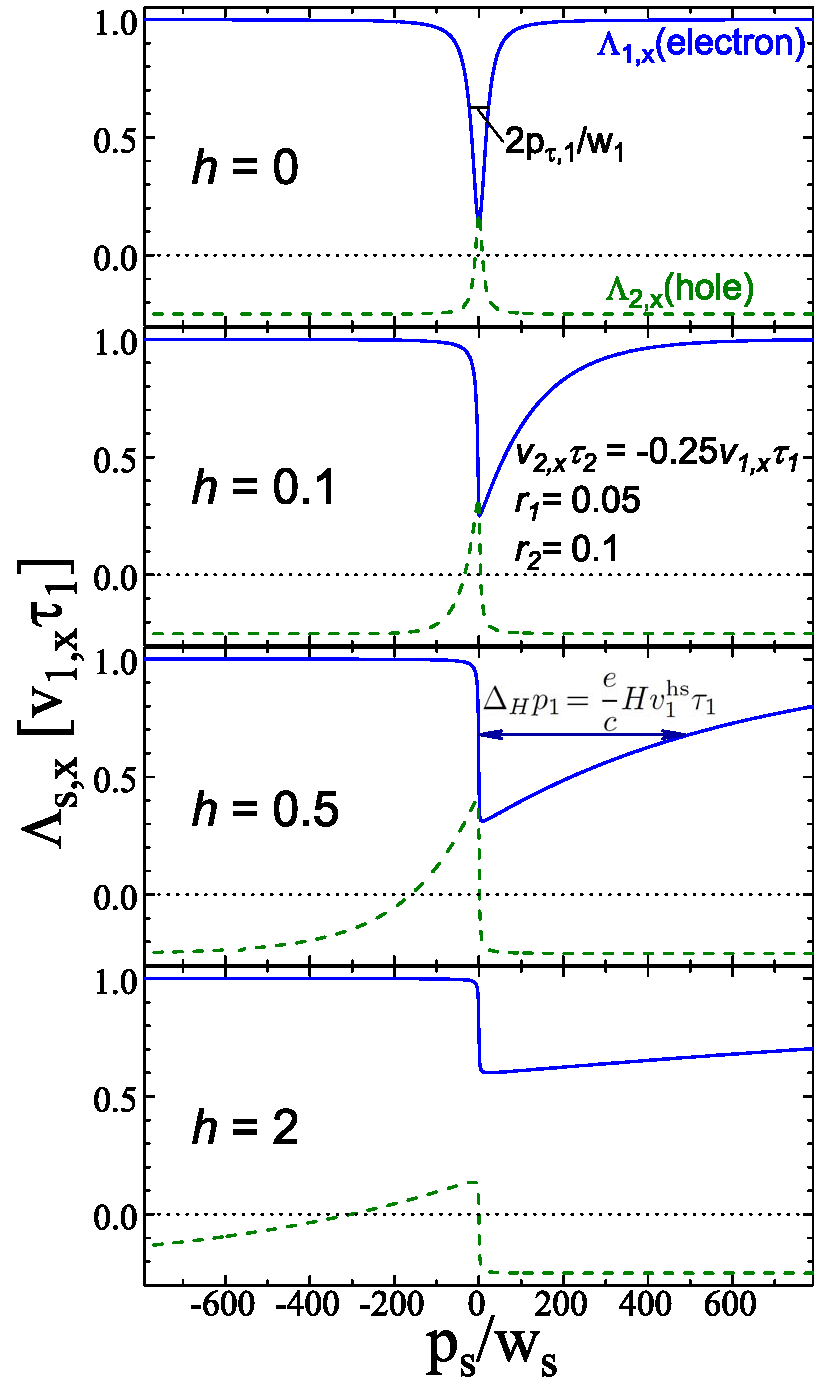}
\caption{The distribution functions $\Lambda_{s,x}(p_{s})$,
Eq.\ (\ref{Lam-ps-H1}), normalized to $v_{1,\alpha}^{\mathrm{hs}}\tau_{1}$
near the hot spot for different magnetic fields and representative parameters
shown in the $h=0.1$ plot. For not too large fields, $h<1$, due to strong
equilibration near hot spot, $\Lambda_{s,x}$ for two bands are very close at
$p_{s}=0$. At zero magnetic field the distribution functions have symmetric
dips with width $\sim p_{\tau,s}$. At fields $h>r_{s}$ the hot spot strongly
disturbs the distribution function within the range $\Delta_{H} p_{s} \propto
H$ on one side ($p_{1}>0$ for the electron band and $p_{2}<0 $ for the hole
band). As a consequence, the dependences $\Lambda_{s,x}(p_{s})$ have steps at
$p_{s}=0$. The height of this step reduces with increasing field for $h>1$. }%
\label{Fig-Lamb-ps-h}%
\end{figure}

Calculation of the conductivity is based on the distribution functions
$\Lambda_{s,x}(p_{s})$ defined by Eq.\ (\ref{Lam-ps-H1}). Figure
\ref{Fig-Lamb-ps-h} illustrates how these functions evolve with increasing
magnetic field. We can see that at zero field the functions have symmetric
dips with width $p_{\tau,s}$, within which they approach almost identical
values at the hot spot, as described in Sec.\ \ref{Sec:CondZeroH}. 
We also note that the hole distribution function changes sign near the hot spot meaning that the partial current
due to the quasiparticles in this region flows in the direction opposite to the average transport current.
This corresponds to the effect of negative transport times caused by strong interband scattering, as
pointed out in Ref.\ \cite{BreitkreizPhysRevB13}. %
At fields $h>r_{s}$ the distribution functions become strongly asymmetric. They are strongly
suppressed at the side from which the hot spot can be reached during the orbital motion in the magnetic
field, i.e., at $p_{1}>0$( $p_{2}<0$) for the electron (hole) band, see Eq.\ (\ref{LamLinear}). The
range of this suppression $\Delta_{H} p_{s}$ grows proportionally to the magnetic field. Due to such
one-side suppression, the distribution functions acquire a steplike features at the hot spot. This
sharp drop reflects small probability of quasiparticle penetration through the hot spot without
scattering. This probability increases with increasing magnetic field and this corresponds to the
step-height decrease.

Using the distribution functions from Eq.\ (\ref{Lam-ps-H1}), we derive from
Eq.\ (\ref{CondLambd}) the magnetic-field dependent part of $S_{\alpha\alpha}(H)$ as
\[
\delta S_{\alpha\alpha}^{\mathrm{hs}}(H) \!=\!-\sum_{s}v_{s,\alpha
}^{\mathrm{hs}}\left(  v_{s,\alpha}^{\mathrm{hs} }\tau_{s}\!-\!\bar{\Lambda
}_{\bar{s},\alpha}\right)  \tau_{s}\gamma_{\mathrm{hs}}\mathcal{F}_{\sigma
}(h,r_{s}).
\]
In Eq.\ (\ref{CondLambd}) we again replaced $v_{s,\alpha}$ by its value at the
hot spot $v_{s,\alpha}^{\mathrm{hs}}$. Substituting $\bar{\Lambda} _{\bar
{s},\alpha}$ from Eq.\ (\ref{LamBarH}), we obtain the field-dependent part of
$S_{\alpha\alpha}$ in the closed form,
\begin{align}
\delta S_{\alpha\alpha}^{\mathrm{hs}}(H)  &  =-\gamma_{\mathrm{hs}}\left(
v_{1,\alpha}^{\mathrm{hs}}\tau_{1}-\!v_{2,\alpha}^{\mathrm{hs}}\tau_{2}\right)
\nonumber\\
\times &  \frac{v_{1,\alpha}^{\mathrm{hs}}\tau_{1}R_{2}\mathcal{F}_{\sigma
,1}-v_{2,\alpha}^{\mathrm{hs}}\tau_{2}R_{1}\mathcal{F}_{\sigma,2}}{R_{1}
+R_{2}-R_{1}R_{2}},\label{SaaH-Result}%
\end{align}
where we used abbreviation $\mathcal{F}_{\sigma,s}\equiv\mathcal{F}_{\sigma
}(h,r_{s})$. This equation determines the field-dependent part of longitudinal
conductivity and represents the main result of this section. At high fields,
$H\gg B_{w,s}$ ($h\gg r_{s}$), this general formula reproduces
Eq.\ (\ref{Saa-HighH}). In the linear regime for $B_{w,s}\ll H \ll B_{\gamma}
$, we can derive somewhat more accurate result, which takes into account a
finite offset,
\[
\delta S_{\alpha\alpha}^{\mathrm{hs}}(H) \approx-\frac{1}{2} \left(  \tau
_{1}v_{1,\alpha}-\!\tau_{2}v_{2,\alpha}\right) ^{2}\frac{e}{c}%
H+S_{\mathrm{off}}%
\]
with $S_{\mathrm{off}}=\frac{\gamma_{\mathrm{hs}}}{2} \left(  \tau
_{1}v_{1,\alpha}-\!\tau_{2}v_{2,\alpha}\right)  \left(  \tau_{1}v_{1,\alpha
}r_{1}-\tau_{2}v_{2,\alpha}r_{2}\right)  $. Note that this offset has the same
order as the zero-field correction, see Eq.\ (\ref{ShszeroH}). At small
fields, $H\ll B_{w,s}$, we obtain
\begin{align*}
\delta S_{\alpha\alpha}^{\mathrm{hs}}(H)  &  \approx-\frac{3\pi^{3/2}}%
{32}\frac{v_{1,\alpha}^{\mathrm{hs}}\tau_{1}-\!v_{2,\alpha}^{\mathrm{hs}}%
\tau_{2}}{\sqrt{\gamma_{\mathrm{hs}}}}\\
\times &  \frac{v_{1,\alpha}^{\mathrm{hs}}|v_{1}^{\mathrm{hs}}|\tau_{1}
^{2}/w_{1}-v_{2,\alpha}^{\mathrm{hs}}|v_{2}^{\mathrm{hs}}|\tau_{2}^{2}/w_{2}%
}{\sqrt{|v_{2}^{\mathrm{hs}}|\tau_{2}/w_{2}}+\sqrt{|v_{1}^{\mathrm{hs}}%
|\tau_{1}/w_{1}}}\left(  \frac{e}{c}H\right)  ^{2}.
\end{align*}
For comparison, the conventional background
contribution\cite{ZimanBook,BlattBook} is given by
\begin{equation}
\delta S_{s,\alpha\alpha}^{(0)}(H) \approx-\tau_{s}^{3}\int\left(
v_{s,\alpha}^{\prime}\right)  ^{2}v_{s}dp_{s}\left(  \frac{e}{c}H\right)  ^{2}%
\end{equation}
with $v_{s,\alpha}^{\prime}=dv_{s,\alpha}/dp_{s}$. We can see that, in
contrast to the zero-field conductivity, the small-field $H^{2}$-correction is
dominated by the hot-spot contribution. It exceeds the background correction
by the factor $\sim p_{F,s}/p_{\tau,s}$.

Equations (\ref{Rs}), (\ref{Fs-Present}), and (\ref{SaaH-Result}) determine
the hot-line contribution to the magnetoconductivity, Eq.\ (\ref{sigma-S}), for
arbitrary values of band Fermi velocities, background scattering rates, and
strength of hot-spot scattering. The qualitative behavior, however, is always
the same: quadratic dependence at very small fields, linear
magnetoconductivity in the intermediate field range, and approaching a
constant value at very high fields. Such behavior was first predicted by Rosch
\cite{RoschPhysRevB00} for a single-band three-dimensional metal near the
antiferromagnetic quantum critical point.

\subsection{Hall conductivity}
\label{Sec:HallCond}

\begin{figure}[ptb]
\includegraphics[width=3.1in]{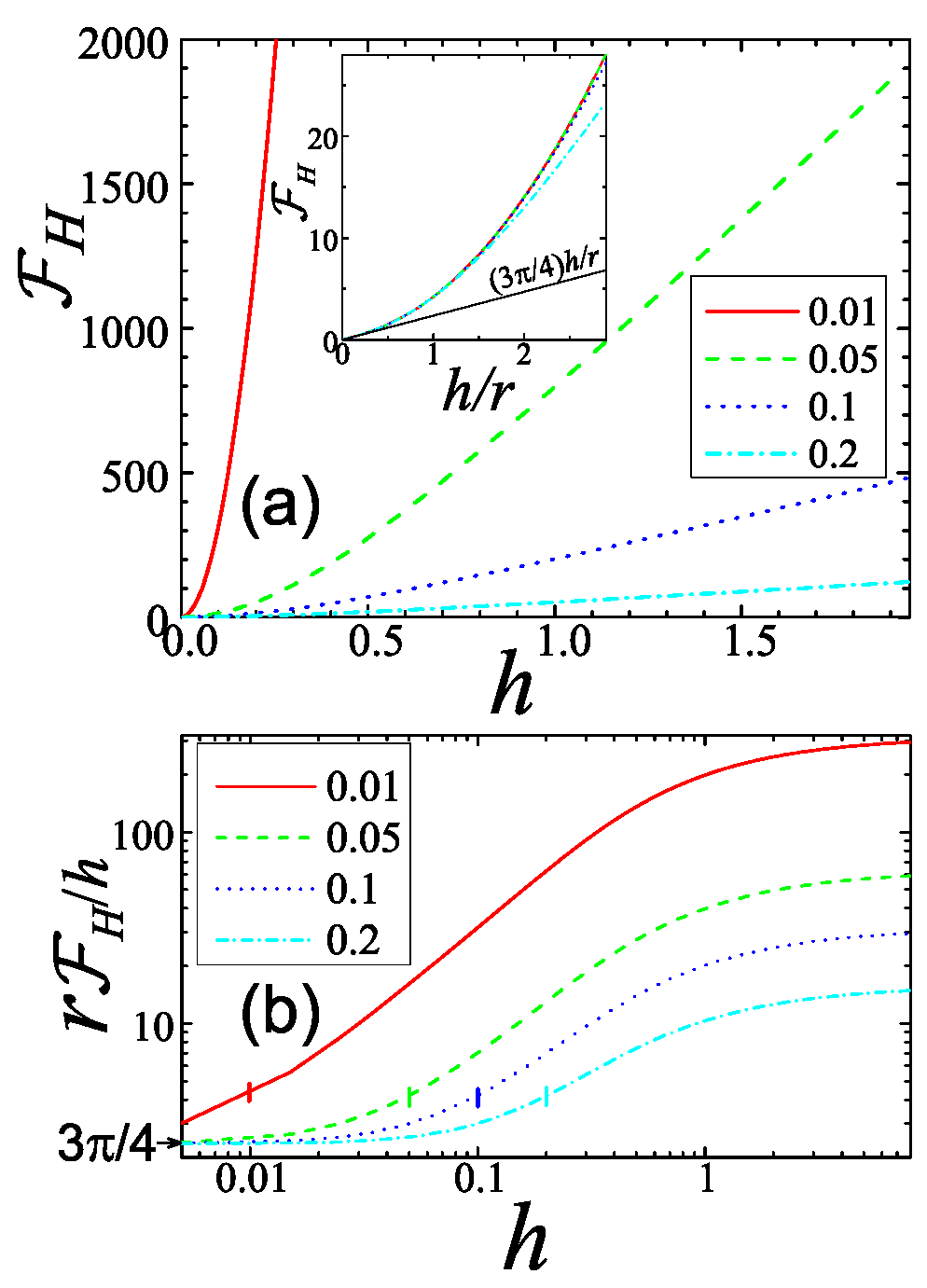}\caption{(a) Plots of the function
$\mathcal{F}_{H}(h,r)$ defined by Eqs.\ (\ref{FHDef}) and (\ref{FHEll}) versus
the reduced field $h$ for different values of $r$ specified in the legend. The
inset shows the plot of $\mathcal{F}_{H}$ vs $h/r$ describing the crossover
between the linear and quadratic regimes at small fields. (b) Double-logarithmic
plots of $r\mathcal{F}_{H}(h,r)/h$ for the same parameters. The vertical bars mark
the values $h=r$ at each plot. These plots clearly illustrate two crossovers
in $\mathcal{F}_{H}(h,r)$ at $h\sim r$ and $h\sim1$.}%
\label{Fig-FH-h}%
\end{figure} %
A finite contribution to the Hall conductivity appears due to the
curvature of the Fermi surface at the hot spot. This means that the dependence
of $v_{s,\alpha}^{\backprime}$ on $p_{s}^{\backprime}$ in
Eqs.\ (\ref{Lam-ps-H1}) and (\ref{Vsa}) can not be neglected. It is sufficient
to keep only the linear-expansion term, $v_{s,\alpha}(p_{s})\approx
v_{s,\alpha}^{\mathrm{hs}}+v_{s,\alpha}^{\prime}p_{s}$ with $v_{s,\alpha
}^{\prime}=dv_{s,\alpha}/dp_{s}$ at $p_{s}=0$. In this approximation the
parameter $\mathcal{V}_{s,\alpha}$, Eq.\ (\ref{Vsa}), can be represented as
\begin{equation}
\mathcal{V}_{s,\alpha}\approx v_{s,\alpha}^{\mathrm{hs}}R_{s}+v_{s,\alpha
}^{\prime}\mathcal{B}_{s}\label{VsaHall}%
\end{equation}
with
\begin{equation}
\mathcal{B}_{s}=\!\frac{1}{\tau_{s}\gamma_{\mathrm{hs}}}\int\limits_{-\infty
}^{\infty}p_{s}^{\backprime}\frac{dp_{s}^{\backprime}}{|v_{s}^{\backprime}
|}\!\int\limits_{-\delta_{s}\infty}^{p_{s}^{\backprime}}\!\frac{dp_{s}}
{\frac{e}{c}H\tau_{s}v_{s}}\mathcal{M}_{H,s}(p_{s}^{\backprime},p_{s}%
),\label{BsDef}%
\end{equation}
where $\mathcal{M}_{H,s}(p_{s}^{\backprime},p_{s})$ is defined by
Eq.\ (\ref{MHsDef}). The field dependence of this function determines behavior
of the Hall conductivity which also has three regimes defined by the field
scales $B_{w,s}$, Eq.\ (\ref{Bw}), and $B_{\gamma}$, Eq.\ (\ref{Bgam}). For
$H>B_{w,s}$ we can again approximate $\psi_{s}(p_{s})$ by $\delta$-function
and this yields the following result
\begin{equation}
\mathcal{B}_{s}\approx\frac{1-\exp\left(  -B_{\gamma}/H\right)  }
{\gamma_{\mathrm{hs}}}\left(  \frac{e}{c}H\right)  ^{2}\tau_{s}v_{s}
^{\mathrm{hs}},\label{BsHighField}%
\end{equation}
meaning that $\mathcal{B}_{s}(H)$ increases \emph{quadratically} with magnetic
field in the range $B_{w,s}<H<B_{\gamma}$ and continues to grow linearly for
$H>B_{\gamma}$. For smaller fields, $H<B_{w,s}$, the dependence $\mathcal{B}%
_{s}(H)$ is sensitive to exact shape of $\psi_{s}(p_{s})$, which we again
assume to be Lorentzian, Eq.\ (\ref{LorentzShape}). In this case, using
Eq.\ (\ref{MHLor}), we can derive the following scaling presentation for
$\mathcal{B}_{s}(H)$,
\begin{equation}
\mathcal{B}_{s}=\delta_{s}w_{s}\mathcal{F}_{H}(H/B_{\gamma},r_{s}
),\label{BsPres}%
\end{equation}
where the reduced function $\mathcal{F}_{H}(h,r)$ is defined by the following
double integral,
\begin{align}
& \mathcal{F}_{H}(h,r) =\frac{r^{2}}{\pi}\int_{0}^{\infty}\!duu\int%
_{0}^{\infty}dz\exp\left(  -z\right) \nonumber\\
& \times\left\{  \exp\left[  -\frac{\arctan\left(  u\!+\!\pi hz/r^{2} \right)
\!-\!\arctan u}{\pi h}\right]  \right. \nonumber\\
& - \left.  \exp\left[  -\frac{\arctan u\!-\!\arctan\left(  u\!-\!\pi
hz/r^{2}\right)  }{\pi h}\right]  \right\} ,\label{FHDef}%
\end{align}
and has the following asymptotics
\[
\mathcal{F}_{H}(h,r)\approx%
\genfrac{\{}{.}{0pt}{}{\frac{3}{4}\pi\frac{h}{r}\left(  1+\frac{5\pi^{3}}
{32}\frac{h^{3}}{r^{3}}\right)  \text{ for }h\ll r}{\pi\frac{h^{2}}{r^{2}
}\left[  1-\exp\left(  -\frac{1}{h}\right)  \right]  \text{ for }h\gg r}%
.
\]
In particular, the last asymptotics reproduces Eq.\ (\ref{BsHighField}). In
the case $r\ll1$ we also derive in the appendix \ref{App:FH} a useful presentation
containing only one integration,
\begin{align}
\mathcal{F}_{H}(h,r)  &  =\frac{h}{r}\int_{0}^{\infty}dxx\exp\left(
-x\right)  G_{H}\left(  \frac{\pi hx}{4r}\right) \nonumber\\
&  +\pi\frac{h^{2}}{r^{2}}\left[  1-\exp\left(  -\frac{1}{h}\right)  \right]
\label{FHEll}%
\end{align}
with
\[
G_{H}(a)\!=\!2\sqrt{a^{2}\!+\!1}E\!\left(\frac{a^{2}}{a^{2}\!+\!1}\right)
\!-\!\frac{1}{2\sqrt{a^{2}\!+\!1}}K\!\left(  \frac{a^{2}}{a^{2}\!+\!1}\right)
\!-\!2a.
\]
Plots of the function $\mathcal{F}_{H}(h,r)$ for different $r$ are shown in
Fig.\ \ref{Fig-FH-h}(a). For clearer illustration of the crossover between
linear and quadratic regimes at small fields, the inset shows plot of
$\mathcal{F}_{H}$ vs $h/r$ for $h<3r$. To demonstrate both crossovers at
$h\sim r$ and $h\sim1$, we show in Fig.\ \ref{Fig-FH-h}(b) the
double-logarithmic plot of $r\mathcal{F}_{H}(h,r)/h$.

We now proceed with derivation of the Hall conductivity. With corrections to
$\mathcal{V}_{s,\alpha}$ given by Eqs.\ (\ref{VsaHall}) and (\ref{BsPres}),
the solution for $\bar{\Lambda}_{s,\alpha}$, Eq.\ (\ref{LamBarH}), becomes
\begin{equation}
\bar{\Lambda}_{s,\alpha}=\sum_{l=s,\bar{s}}\left(  U_{s,l}v_{l,\alpha}\tau
_{l}+W_{s,l}\delta_{l}w_{l}v^{\prime}_{l,\alpha}\tau_{l}\right)
\label{LamBarHall}%
\end{equation}
with
\begin{align*}
U_{s,s} & =\frac{R_{s}}{R_{1}\!+\!R_{2}\!-\!R_{1}R_{2}},\ U_{s,\bar{s}}%
=\frac{R_{\bar{s}}\left(  1-R_{s}\right) }{R_{1}\!+\!R_{2}\!-\!R_{1}R_{2}},\\
W_{s,s} & =\frac{\mathcal{F}_{H,s}}{R_{1}\!+\!R_{2}\!-\!R_{1}R_{2}%
},\ W_{s,\bar{s}}=\frac{\mathcal{F}_{H,\bar{s}}\left(  1-R_{s}\right) }%
{R_{1}\!+\!R_{2}\!-\!R_{1}R_{2}},%
\end{align*}
where we introduced abbreviations $\mathcal{F}_{H,s} \equiv\mathcal{F}%
_{H}(H/B_{\gamma},r_{s})$. From Eqs.\ (\ref{CondLambd}) and (\ref{Lam-ps-H1})
we find that the Hall conductivity is determined by $S_{xy}=\sum_{s}S_{s,xy}$
with
\begin{align}
S_{s,xy}  &  \approx\int\frac{dp_{s}}{|v_{s}|}v_{s,x}\nonumber\\
\times &  \left(  \bar{\Lambda}_{\bar{s},y}+ \int\limits_{p_{s}}^{\delta
_{s}\infty}\!dp_{s}^{\backprime} \frac{\tau_{s}v_{s,y}^{\backprime}%
\!-\!\bar{\Lambda}_{\bar{s},y}} {\frac{e}{c}Hv_{s}^{\backprime}\tau_{s}}
\mathcal{L}_{H,s}(p_{s}^{\backprime},p_{s})\right) ,
\end{align}
where $\mathcal{L}_{H,s}(p_{s}^{\backprime},p_{s})$ is defined by
Eq.~(\ref{LHsDef}) and for the Lorentzian hot spot can be estimated as
\begin{align}
\mathcal{L}_{H,s}(p_{s}^{\backprime},p_{s})  &  \approx\exp\left[
-\frac{p_{s}^{\backprime}\!-\!p_{s}}{\frac{e}{c}Hv_{s}\tau_{s}}\right.
\nonumber\\
&  -\left.  \frac{\gamma_{\mathrm{hs}}}{\frac{e}{c}\pi H}\left(  \arctan
\frac{p_{s}^{\backprime}}{w_{s}}\!-\!\arctan\frac{p_{s}}{w_{s}}\right)
\right] .\label{LHLor}%
\end{align}
First, we separate from $S_{s,xy}$ a conventional background contribution,
\begin{align}
S_{s,xy}^{(0)}  &  =\int\frac{dp_{s}}{|v_{s}|}v_{s,x} \int\limits_{p_{s}%
}^{\delta_{s}\infty}\ \frac{dp_{s}^{\backprime}}{v_{s}^{\backprime}}
\frac{v_{s,y}^{\backprime}}{\frac{e}{c}H}\exp\left[  -\frac{p_{s}^{\backprime
}-p_{s}}{\frac{e}{c}Hv_{s}\tau_{s}}\right] \nonumber\\
&  =-\delta_{s}\frac{e}{c}H\tau_{s}^{2}\int dp_{s}v_{s,x}v_{s,y}^{\prime
}.\label{Sxy0}%
\end{align}
Subtracting this term, we obtain the hot-spot contribution, $S_{s,xy}%
^{\mathrm{hs}}=S_{s,xy}-S_{s,xy}^{(0)}$, as
\begin{equation}
S_{s,xy}^{\mathrm{hs}}\!\approx\!\int\frac{dp_{s}}{|v_{s}|}v_{s,x}
\!\int\limits_{p_{s}}^{\delta_{s}\infty}\!\frac{dp_{s}^{\backprime}} {\frac
{e}{c}Hv_{s}^{\backprime}\tau_{s}} \left(  \bar{\Lambda}_{\bar{s},y}%
\!-\!\tau_{s}v_{s,y}^{\backprime}\right)  \mathcal{M}_{H,s}(p_{s}^{\backprime
},p_{s}),\label{Ss-hs}%
\end{equation}
where $\mathcal{M}_{H,s}(p_{s}^{\backprime},p_{s})$ is defined by
Eq.\ (\ref{MHLor}). Substituting $\bar{\Lambda}_{s,\alpha}$ from
Eq.\ (\ref{LamBarHall}) and expanding $v_{s,y}$ and $v_{s,x}^{\backprime}$
near the hot spot, after some algebraic transformations, we finally find the
total hot-spot Hall term
\begin{align}
S_{xy}^{\mathrm{hs}}  &  =-\frac{\gamma_{\mathrm{hs}}}{R_{1}+R_{2}-R_{1}R_{2}
}\sum_{s}\delta_{s}w_{s}\mathcal{F}_{H,s}R_{\bar{s}}\mathcal{K}_{s}
,\label{SxyResult}\\
\mathcal{K}_{s}  &  =\tau_{s}\left(  \tau_{s} \left[ \mathbf{v}_{s}%
^{\mathrm{hs}}\times\mathbf{v}_{s}^{\prime}\right] _{z} \!+\!\tau_{\bar{s}%
}\left[  \mathbf{v}_{s}^{\prime}\times\mathbf{v}_{\bar{s}}^{\mathrm{hs}%
}\right] _{z} \right) ,\nonumber
\end{align}
which determines the Hall conductivity via Eq.\ (\ref{sigma-S}). We can see
that, in general, $S_{xy}^{\mathrm{hs}}$ contains both intraband and interband contributions.

As follows from Eq.\ (\ref{SxyResult}), the hot-spot Hall conductivity has
three asymptotic regimes: (i) Small-field linear regime, $h\ll r_{s}$,
\begin{align}
S_{xy}^{\mathrm{hs}} &  \approx-\frac{3}{4}\pi\frac{\gamma_{\mathrm{hs}}%
h}{r_{1}+r_{2}}\sum_{s}\delta_{s}w_{s}\frac{r_{\bar{s}}}{r_{s}}\mathcal{K}%
_{s}\nonumber\\
&  =\frac{3}{4}\frac{\frac{e}{c}H\sqrt{\pi\gamma_{\mathrm{hs}}}}{\sqrt
{|v_{1}^{\mathrm{hs}}|\tau_{1}/w_{1}}+\sqrt{|v_{2}^{\mathrm{hs}}|\tau
_{2}/w_{2}}}\sum_{s}v_{s}^{\mathrm{hs}}\tau_{s}\mathcal{K}_{s}%
,\label{Sxy-smallH}%
\end{align}
(ii) Intermediate quadratic regime, $r_{s}\ll h\ll1$,
\begin{align}
S_{xy}^{\mathrm{hs}} &  \approx-\frac{\pi}{2}\gamma_{\mathrm{hs}}h^{2}\sum
_{s}\frac{\delta_{s}w_{s}}{r_{s}^{2}}\mathcal{K}_{s}\nonumber\\
&  =-\frac{1}{2}\left(  \frac{e}{c}H\right)  ^{2}\sum_{s}v_{s}^{\mathrm{hs}%
}\tau_{s}\mathcal{K}_{s},\label{Sxy-intermH}%
\end{align}
and (iii) Large-field linear regime, $h>1$,
\begin{align}
S_{xy}^{\mathrm{hs}} &  \approx-\pi\gamma_{\mathrm{hs}}h\sum_{s}\frac
{\delta_{s}w_{s}}{r_{s}^{2}}\mathcal{K}_{s}\nonumber\\
&  =\gamma_{\mathrm{hs}}\frac{e}{c}H\sum_{s}v_{s}^{\mathrm{hs}}\tau
_{s}\mathcal{K}_{s}.\label{Sxy-highH}%
\end{align}
The latter two asympotics correspond to Eq.\ (\ref{BsHighField}). Note that in
all three asymptotics $S_{xy}^{\mathrm{hs}}$ is proportional to $\sum_{s}%
v_{s}^{\mathrm{hs}}\tau_{s}\mathcal{K}_{s}$ even though the general result,
Eq.\ (\ref{SxyResult}), does not have this property. Comparing $S_{xy}%
^{\mathrm{hs}}$, Eq.\ (\ref{SxyResult}), and its asymptotics with the
conventional contribution $S_{xy}^{(0)}$, Eq.\ (\ref{Sxy0}), we can make
several observations. The signs of the hot-spot correction terms are opposite
to the corresponding conventional contributions. The correction to the linear
Hall conductivity at $H<B_{w,s}$ is typically small. The relative correction
is of the order of $p_{\tau}/p_{F}$, similar to the zero-field conductivity.
However, the hot-spot correction leads to crossover to quadratic regime at
relatively small magnetic fields, $H\sim B_{w,s}$, and this quadratic field
dependence persists within a wide range of the magnetic fields. In combination with
the linear decrease of the longitudinal conductivity, this behavior provides
clear signatures of the hot-spot scattering. Comparing Eqs.\ (\ref{Sxy0}) and
(\ref{Sxy-highH}), we can see that the overall relative change of slope of the
partial Hall conductivity for bands $s$, $\sigma_{xy,s}$, from very small to
very large field is determined by the hot-spot strength as
\[
\frac{\Delta\sigma_{xy,s}^{\prime}}{\sigma_{xy,s}^{\prime}}\sim\frac
{\gamma_{\mathrm{hs}}|v_{s}|\tau_{s}}{p_{F,s}}%
\]
with $\sigma_{xy,s}^{\prime}=\partial\sigma_{xy,s}/\partial H$ and
$\Delta\sigma_{xy,s}^{\prime}=\sigma_{xy,s}^{\prime}(H\gg B_{\gamma}%
)-\sigma_{xy,s}^{\prime}(H\rightarrow0)$.

\section{Representative magnetic field dependences for a simple four-band model}
\label{Sec-MagFieldDep} 

\begin{figure*}[ptb]
\includegraphics[width=5.7in]{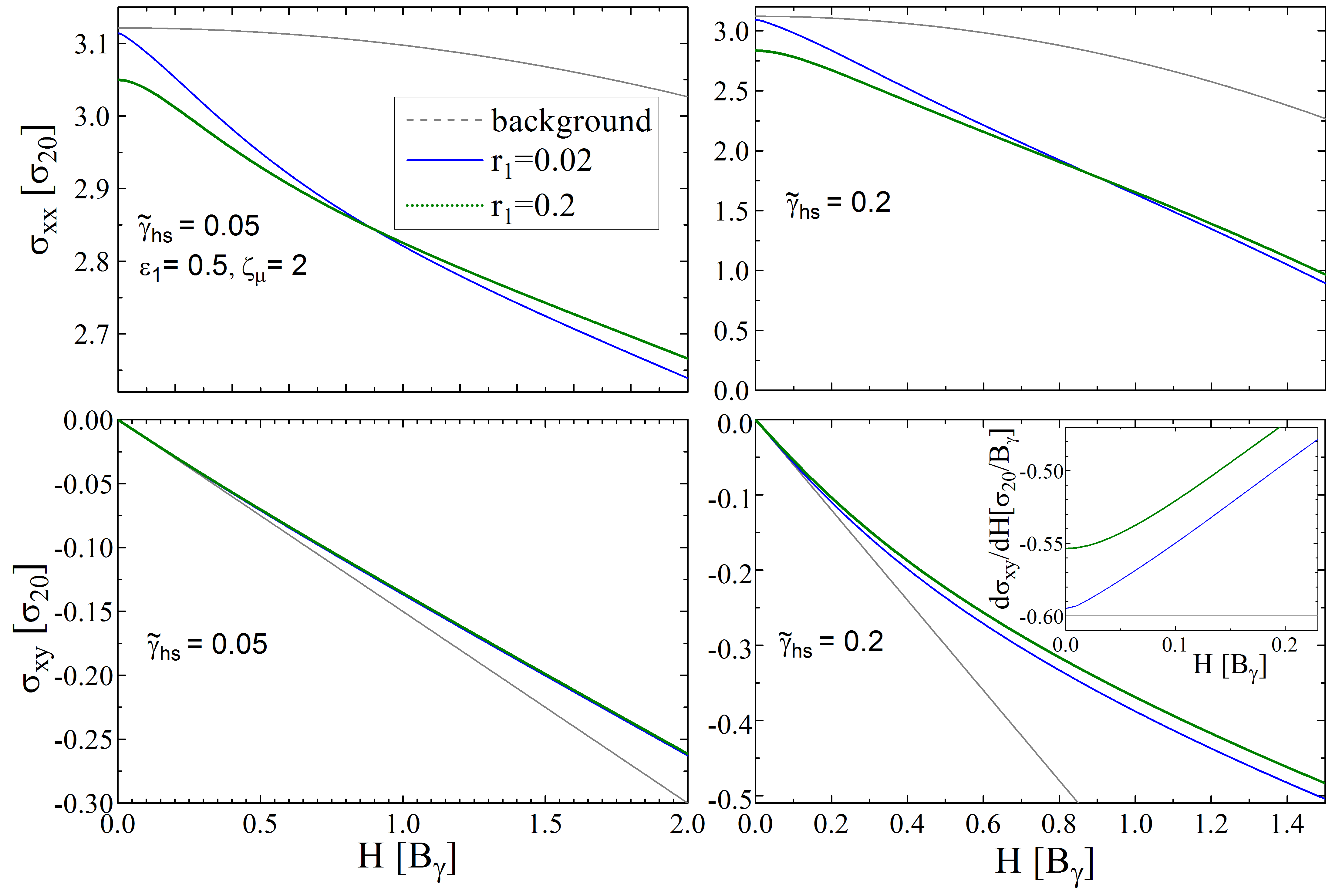}
\caption{Representative magnetic field
dependences of the conductivity components for two values of hot-spot strength
$\tilde{\gamma}_{\mathrm{hs}}$, $0.05$ (left column) and $0.2$ (right column)
and two values of ``sharpness'' parameter $r_{1}$, $0.02$ and $0.2$ (solid and
dotted lines). For reference, we also show background conductivity without
hot-spot scattering (dashed lines). The inset in the lower right plot show the
derivative $d\sigma_{xy}/dH$ at low fields to emphasize the difference between
two values of $r_{1}$.}%
\label{Fig-Sigma-H}%
\end{figure*}
In this section we illustrate general trends in the magnetic-field dependences
of the conductivity components for different parameters. The iron pnictides and chalcogenides
typically have at least two hole bands in the Brillouin zone center and two
electron bands at the zone edge. This makes fully realistic analysis rather
complicated and requires knowledge of many band-structure and scattering
parameters. For illustration, we consider a minimum model of compensated metal
with two identical hole and two electron Fermi surfaces. We assume circular
and elliptical cross sections for the hole and electron Fermi surfaces
respectively,
\begin{equation}
\xi_{1,\mathbf{p}}=\varepsilon_{1,0}-\mu+\frac{p_{x}^{2}}{2m_{x}}+\frac
{p_{y}^{2}}{2m_{y}},\ \xi_{2,\mathbf{p}}=\varepsilon_{2,0}-\mu-\frac{p^{2}
}{2m_{2}},\nonumber
\end{equation}
which are characterized by the Fermi momenta, $p_{F,2}=\sqrt{2m_{2}
\varepsilon_{F,2}}$ and $p_{F,\alpha}=\sqrt{2m_{\alpha}\varepsilon_{F,1}}$
with $\alpha=x,y$, $\varepsilon_{F,1}=\mu-\varepsilon_{1,0}$, $\varepsilon
_{F,2}=\varepsilon_{2,0}-\mu$. In the further analysis, we will assume that
the inequality $p_{F,x}>p_{F,2}>p_{F,y}$ holds. The second electron band is
90$^{\circ}$-rotated with respect to the first one. Each electron band has
four hot spots and each hole band has eight hot spots. Introducing the ratios,
$u_{\alpha}=p_{F,2}/p_{F,\alpha}$ with $u_{x}<1$ and $u_{y}>1$, we find the
cosine and sine of the hot-spot angle $\theta_{\mathrm{hs}}$ as%
\begin{equation}
\cos^{2}\theta_{\mathrm{hs}}=\frac{u_{y}^{2}-1}{u_{y}^{2}-u_{x}^{2}}%
,\ \sin^{2}\theta_{\mathrm{hs}}=\frac{1-u_{x}^{2}}{u_{y}^{2}-u_{x}^{2}%
}.\label{CosSinHS}%
\end{equation}
For the compensated case $u_{x}u_{y}=1$.

In the previous sections we focused on a single $p_{z}$ cross section of the
Fermi surface. Calculation of the conductivity in Eq.\ (\ref{sigma-S})
includes the integration over $p_{z}$, which means averaging over all cross
sections. For estimate, we will use result for a single representative cross
section. As the conductivity unit, we take the partial conductivity of
the hole bands at zero magnetic field, $\sigma_{20}\equiv\sigma_{2,xx}%
(0)\propto S_{2,xx}^{(0)}(0)$. We also introduce notations for the in-plane
mass anisotropy of the electron band $\epsilon_{1}=m_{y}/m_{x}$, the average
mobility ratio $\zeta_{\mu}=m_{2}\tau_{1}/\bar{m}\tau_{2}$ with $\bar{m}%
=\sqrt{m_{x}m_{y}}$, and the reduced hot-spot strength%
\[
\tilde{\gamma}_{\mathrm{hs}}=\gamma_{\mathrm{hs}}|v_{2}|\tau_{2}/p_{F,2}\ll1.
\]
In these notations the ratios of the mobility components are $\tau_{1}%
v_{F,x}/\tau_{2}|v_{2}|=\xi_{\mu}/u_{y}$, $\tau_{1}v_{F,y}/\tau_{2}|v_{2}%
|=\xi_{\mu}/u_{x}$.

Using the reduced parameters, we obtain the following presentations for the
background zero-field conductivity
\[
\sigma_{xx}^{(0)}(0)=\sigma_{20}\left[  1+\zeta_{\mu}\left(  u_{x}^{-2}%
+u_{y}^{-2}\right)  /2\right],
\]
magnetoconductivity
\[
\delta\sigma_{xx}^{(0)}(h)=-\sigma_{20}\tilde{\gamma}_{\mathrm{hs}}^{2}%
h^{2}\left[  1+\left(  u_{x}^{-2}+u_{y}^{-2}\right)  \zeta_{\mu}^{3}/2\right]
,
\]
and Hall conductivity%
\[
\sigma_{xy}^{(0)}=\sigma_{20}\tilde{\gamma}_{\mathrm{hs}}h\left(
1-\frac{\zeta_{\mu}^{2}}{u_{x}u_{y}}\right)  .
\]
Note that the parameter $\tilde{\gamma}_{\mathrm{hs}}$ appears in these
presentation only because we use the field scale $B_{\gamma}$ which is
proportional to $\gamma_{\mathrm{hs}}$.

The contributions from the four hot spots are determined by the ratios of
the local mobilities at these points which can be evaluated as $\tau_{1}%
v_{1,x}^{\mathrm{hs}}/\tau_{2}|v_{2,x}^{\mathrm{hs}}|\!=\!\xi_{\mu}%
\sqrt{\epsilon_{1}} $ and $\tau_{1}v_{1,y}^{\mathrm{hs}}/\tau_{2}%
|v_{2,y}^{\mathrm{hs}}|$ $=\!\xi_{\mu}/\sqrt{\epsilon_{1}}$. We also obtain
relation between the parameters $r_{s}$, Eq.\ (\ref{rsLorentz}), $r_{2}%
^{2}=r_{1}^{2}\frac{w_{2}}{w_{1}}\zeta_{\mu}\sqrt{u_{x}^{-2}+u_{y} ^{-2}-1}$
and assume that $w_{1}=w_{2}$. The hot-spot contributions to the zero-field
conductivity, Eq.\ (\ref{ShszeroH}) and longitudinal magnetoconductivity,
Eq.\ (\ref{SaaH-Result}), can now be presented as
\begin{align}
&  \sigma_{xx}^{\mathrm{hs}}(0)=-\frac{4}{\pi}\frac{\sigma_{20}\tilde{\gamma
}_{\mathrm{hs}}}{1/r_{1}+1/r_{2}}\nonumber\\
&  \times\!\left[  \cos^{2}\theta_{\mathrm{hs}}\left(1\!+\zeta_{\mu}
\sqrt{\epsilon_{1}}\right)  ^{2}\!+\sin^{2}\theta_{\mathrm{hs}}\left(
1\!+\zeta_{\mu}/\sqrt{\epsilon_{1}}\right)^{2}  \right]  ,\label{SHSZero-R}%
\end{align}
and
\begin{align}
&  \delta\sigma_{xx}^{\mathrm{hs}}(H)=-\frac{4}{\pi}\frac{\sigma_{20}
\tilde{\gamma}_{\mathrm{hs}}}{R_{1}+R_{2}-R_{1}R_{2}}\nonumber\\
&  \times\Big[\cos^{2}\theta_{\mathrm{hs}}\left(  \zeta_{\mu}\sqrt
{\epsilon_{1}}+\!1\right)  \left(  \zeta_{\mu}\sqrt{\epsilon_{1}}%
R_{2}\mathcal{F}_{\sigma,1}+R_{1}\mathcal{F}_{\sigma,2}\right) \nonumber\\
&  +\sin^{2}\theta_{\mathrm{hs}}\left(  \frac{\zeta_{\mu}}{\sqrt{\epsilon_{1}
}}+\!1\right)  \left(  \frac{\zeta_{\mu}}{\sqrt{\epsilon_{1}}}R_{2}
\mathcal{F}_{\sigma,1}+R_{1}\mathcal{F}_{\sigma,2}\right)
\Big],\label{dSHSxx-R}%
\end{align}
respectively. For the derivation of the Hall term, Eq.\ (\ref{SxyResult}), we
obtain the following relations
\[
\frac{\tau_{1}\left[  \mathbf{v}_{1}^{\mathrm{hs}}\times\mathbf{v}_{1}
^{\prime}\right]  _{z}}{\tau_{2}\left[  \mathbf{v}_{1}^{\prime}\times
\mathbf{v}_{2}^{\mathrm{hs}}\right]  _{z}}=\frac{\tau_{1}\left[
\mathbf{v}_{2}^{\prime}\times\mathbf{v}_{1}^{\mathrm{hs}}\right]  _{z}}%
{\tau_{2}\left[  \mathbf{v}_{2}^{\mathrm{hs}}\times\mathbf{v}_{2}^{\prime
}\right]  _{z}}=\frac{\tau_{1}\varepsilon_{F,1}}{\tau_{2}\varepsilon_{F,2}%
}=\frac{\xi_{\mu}}{u_{x}u_{y}},
\]
which allow us to present this term as
\begin{equation}
\sigma_{xy}^{\mathrm{hs}}(H)\!=\!\frac{8}{\pi^{2}}\sigma_{20} \tilde{\gamma
}_{\mathrm{hs}}^{2}\frac{r_{1}^{2}\zeta_{\mu}^{2}\mathcal{F}_{H,1}
R_{2}\!-\!r_{2}^{2}\mathcal{F}_{H,2}R_{1}}{R_{1}+R_{2}-R_{1}R_{2}}\left(
1\!+\!\frac{\zeta_{\mu}}{u_{x}u_{y}}\right)  .\label{SHSxy-R}%
\end{equation}
We utilize the derived presentations for illustration of the possibles shapes
of $\sigma_{\alpha\beta}(H)$ dependences.

The overall behavior of the conductivity components mostly depends on the two reduced
parameters $\tilde{\gamma}_{\mathrm{hs}}$ and $r_{1}$. The first parameter
determines the magnitude of the hot-spot correction with respect to
background while the second parameter determines the behavior at small
magnetic fields. Figure \ref{Fig-Sigma-H} shows the representative magnetic-field
dependences of the conductivity components for two values of $\tilde{\gamma
}_{\mathrm{hs}}$, $0.05$ and $0.2$ and two values of $r_{1}$, $0.02$ and
$0.2$. As a reference, we also show the background longitudinal and Hall
conductivities without hot-spot scattering.

We can see that for the weak hot spot, $\tilde{\gamma}_{\mathrm{hs}}=0.05$, the
corrections are small while for $\tilde{\gamma}_{\mathrm{hs}}=0.2$ they become comparable
with the background. In particular, the slope of the Hall conductivity $|d\sigma_{xy}/dH|$
drops more than twice with increasing magnetic field. The role of the ``sharpness''
parameter $r_{1}$ is more obvious for the longitudinal conductivity. For the broad hot
spot, $r_{1}=0.2$, we can see the region of quadratic magnetoconductivy for $h<0.2$. For
the narrow hot spot, $r_{1}=0.02$, this region is practically invisible in the plots and
the conductivity has linear magnetic-field dependence in the extended field range. In
contrast, for $r_{1}=0.2$ this linear regime is not pronounced and looks more like an
inflection point. The parameter $r_{1}$ only weakly influences the shape of
$\sigma_{xy}(H)$ because it mostly determines the small correction to the low-field linear
slope. This small correction can be more clearly seen in the field dependence of the
derivative $d\sigma_{xy}/dH$ at small fields, see the inset in the lower right plot of
Fig.\ (\ref{Fig-Sigma-H}). Also, the inset plots clearly demonstrate that the hot-spot
correction to $\sigma_{xy}$ has quadratic magnetic-field dependence in the intermediate
field range.

\section{Summary and discussion}
\label{Sec:Discussion}

In summary, we analyzed in detail magnetotransport due to the hot-spot
scattering on the AF fluctuations in multiple-band metals. The key qualitative
features are extended ranges of the linear magnetic-field dependence of
longitudinal conductivity and, simultaneously, the quadratic dependence of the
Hall component.

This mechanism is very likely responsible for anomalous magnetotransport
properties found in some iron pnictides and chalcogenides in paramagnetic state.
For example, the linear magnetoresistance and strongly nonlinear Hall
resistivity have been found in Fe$_{1+y}$Te$_{0.6}$Se$_{0.4}$%
\cite{SunPhysRevB14}. This behavior becomes pronounced after annealing which
strongly reduces background scattering. Such behavior roughly corresponds to
illustration in Fig.\ \ref{Fig-Sigma-H} for strong and narrow hot spot,
$\tilde{\gamma}_{\mathrm{hs}}\!=\!0.2$ and $r_{1}\!=\!0.02$ for $H<B_{\gamma}%
$. The absence of saturation at high magnetic fields simply means that the
field scale $B_{\gamma}$ for this compound is very high, more than $30$ tesla.
In other compound, Ba[As$_{1-x}$P$_{x}$]$_{2}$Fe$_{2}$, longitudinal
resistance has small but clear deviations from quadratic magnetic-filed
dependence \cite{WelpUnpubl,AnalytisNPhys14} while the Hall resistance has
weakly nonlinear field dependence \cite{WelpUnpubl}. Such behavior resembles
illustration in Fig.\ \ref{Fig-Sigma-H} for weak and broad hot-spot,
$\tilde{\gamma}_{\mathrm{hs}}\!=\!0.05$ and $r_{1}\!=\!0.2$. The corresponding
typical magnetic fields seem to be rather large, $B_{w} \sim35$T and
$B_{\gamma}\sim65$T. Anomalous properties are seen in the optimally-doped compound
and they disappear in overdoped compounds. This is consistent with
the interpretation based on the spin-fluctuation scattering. In principle, detailed
analysis of magnetotransport allows to extract the hot-spot parameters which
would give us a valuable microscopic information about properties of spin fluctuations.

It is instructive to compare  behavior of magnetotransport  in the paramagnetic state due
to hot-spot scattering and in the antiferromagnetic state due to Fermi surface reconstruction
\cite{FentonPRL05,LinPRB05,KoshelevPRB13}. In both cases the anomalous behavior is caused
by the interruption of smooth orbital motion of quasiparticles along the Fermi surface in
the magnetic field.  In the first case the interruption is caused by the sharp enhancement of 
scattering and in the second case by the abrupt change of the Fermi velocity. In both cases
there is a low-field crossover at which the field dependence of the longitudinal conductivity
changes from quadratic to linear while the dependence of the Hall conductivity changes
from linear to quadratic.   For the hot-spot mechanism the crossover field is determined
by the scattering strength and width of the hot spots and for the reconstruction mechanism
it is determined be the antiferromagnetic gap. Above the crossover both the hot spots and
turning points can be treated as sharp regions  and the conductivity components are not
sensitive to their internal structure. We can note that for identical background
scattering times $\tau_1=\tau_2$ in the linear regime  the longitudinal conductivity due
to the hot-spots is two times smaller than one due to the reconstruction mechanism. For
the hot-spot mechanism the linear (quadratic) growth of the longitudinal (Hall)
conductivity is limited from above by the second magnetic field scale determined by the total
scattering strength. Such limit is absent for the reconstruction mechanism.

\begin{acknowledgements}
	
This study was initiated by the puzzling high-field magnetotransport data of the optimally-doped
compound Ba[As$_{1-x}$P$_{x}$]$_{2}$Fe$_{2}$ \cite{WelpUnpubl} (the behavior of the
longitudinal conductivity is very similar to reported in Ref.\
\onlinecite{AnalytisNPhys14}). The author would like to thank U.\ Welp and Y.\ Jia for
useful discussion of these data. This work was supported by the U.S. Department of Energy,
Office of Science, Materials Sciences and Engineering Division.
\end{acknowledgements}

\appendix

\section{Function $\mathcal{F}_{\sigma}(h,r)$ for $r\ll 1$}

\label{App-Fs-small-rApp}

In the case $r\ll1$ the two crossovers in the behavior of the function
$\mathcal{F}_{\sigma}(h,r)$ , Eq.\ (\ref{Fs-h-r}), are well separated. This
allows us to derive a useful presentation for this function containing only
single integration. In the region $h\ll1$ the ratio $\mathcal{F}_{\sigma
}(h,r)/r$ depends on the single parameter $h/r$. Therefore, for the analysis of this
region, it is convenient to introduce new function $F_{\sigma}(b,r)$ defined
by the relation
\begin{equation}
\mathcal{F}_{\sigma}(h,r)=(r/\pi)F_{\sigma}(\pi h/r,r).\label{Fs-redefine}%
\end{equation}
This new function is defined as 
\begin{widetext}
	\begin{equation}
		F_{\sigma}(b,r)=\int_{-\infty}^{\infty}dv\int_{0}^{\infty}dz\exp\left(
		-z\right)  \left\{  \exp\left[  -\frac{z}{r^{2}+v^{2}}\right]  -\exp\left[
		-\frac{\arctan\left(  \left(  v+\frac{bz}{2}\right)  /r\right)  -\arctan
			\left(  \left(  v-\frac{bz}{2}\right)  /r\right)  }{rb}\right]  \right\},
		\label{Fs-b-r}
	\end{equation}
where $b=\pi h/r$ is the redefined reduced field. For $r\ll1,\pi/b$, using asympotics
$\arctan x\approx \pm \pi/2 \mp 1/x$ for $x\rightarrow \pm\infty$, we can approximate
\[
	\arctan\left(  \frac{v+\frac{bz}{2}}{r}\right)  -\arctan\left(  \frac
	{v-\frac{bz}{2}}{r}\right)  \approx
	\genfrac{\{}{.}{0pt}{}{\frac{rbz}{v^{2}-\frac{\left(  bz\right)  ^{2}}{4}
		}\;\text{, for}\;|v|-\frac{bz}{2}\gg r}{\pi-\frac{2vr}{v^{2}-\frac{\left(
			bz\right)  ^{2}}{4}}\;\text{, for}\;\frac{bz}{2}-|v|\gg r}.
\]
\end{widetext}
We can conclude that the region $|v|<\frac{bz}{2}$ gives very small
contribution to the second-term integration in Eq.\ (\ref{Fs-b-r}) because it
contains exponentially small factor $\exp(-\pi/rb)$. Therefore, we can rewrite
$F_{\sigma}(b,r)$ as
\begin{align*}
F_{\sigma}(b) & =\int_{0}^{\infty}dz\exp\left(  -z\right)  \left[
\int_{|v|>0}dv\exp\left(  -\frac{z}{v^{2}}\right)  \right. \\
& - \left. \int_{|v|>\frac{bz}{2}} dv\exp\left(  -\frac{z}{v^{2}-\frac{\left(
bz\right)  ^{2}}{4}}\right)  \right].
\end{align*}
We can see that this function does not depend explicitly on the parameter $r$.
It describes the crossover between the quadratic and linear regimes for
$b\ll\pi/r$. Performing variable change $v\rightarrow v+bz/2$ in the second
term, we obtain
\begin{align*}
&  \int_{0}^{v_{0}}dv\exp\left(  -\frac{z}{v^{2}}\right)  -\int_{\frac{bz}{2}
}^{v_{0}}dv\exp\left(  -\frac{z}{v^{2}-\frac{\left(  bz\right)  ^{2}}{4}
}\right) \\
&  =\int_{0}^{v_{0}}dv\exp\left(  -\frac{z}{v^{2}}\right)  -\int_{0}
^{v_{0}-\frac{bz}{2}}dv\exp\left(  -\frac{z}{v^{2}+bzv}\right) \\
&  =\int_{0}^{v_{0}}dv\left[  \exp\left(  -\frac{z}{v^{2}}\right)
-\exp\left(  -\frac{z}{v^{2}+bzv}\right)  \right]  +\frac{bz}{2},
\end{align*}
where $v_{0}\gg 1$ is an arbitrary upper cut off which can be sent to infinity in the
last formula. After these transformations, $F_{\sigma}(b)$ takes the following
form
\begin{align*}
F_{\sigma}(b)  &  =2\int\limits_{0}^{\infty}dz\int\limits_{0}^{\infty}
dv\exp\left(  -z\right) \\
\times &  \left[  \exp\left(  -\frac{z}{v^{2}}\right)  -\exp\left(  -\frac
{z}{v^{2}+bzv}\right)  \right]  +b\\
=2  &  \int\limits_{0}^{\infty}dv\left[  \frac{v^{2}}{v^{2}+1}-\int%
_{0}^{\infty}dz\exp\left(  -z-\frac{z}{v^{2}+bzv}\right)  \right]  +b.
\end{align*}
To eliminate a complicated expression in the exponent, we make the variable
change
\begin{align*}
&  x=z+\frac{z}{v^{2}+bzv},\\
z  &  =-\frac{\left(  v^{2}+1-xbv\right)  }{2vb}+\sqrt{\frac{\left(
v^{2}+1-xbv\right)  ^{2}}{4v^{2}b^{2}}+\frac{xv}{b}},
\end{align*}
which allows us to present $F_{\sigma}(b)$ as
\begin{equation}
F_{\sigma}(b)=\int_{0}^{\infty}dx\exp\left(  -x\right)  G\left(  \frac{bx}
{4}\right)  -\pi\label{Fs-Ga}%
\end{equation}
with
\[
G(a)=\int_{0}^{\infty}dv\left(  1-\frac{v^{2}-1+4av}
{\sqrt{\left(v^{2}\!-1+4av\right)  ^{2}+4v^{2}}}\right)  +4a.
\]
Introducing a new variable $\zeta$ defined by $2\zeta=v-1/v+2a$ with the
inverse relation $v=\left(  \zeta-a\right)  +\sqrt{\left(  \zeta-a\right)
^{2}+1}$, we obtain the presentation
\begin{equation}
G(a)=2\int_{0}^{\infty}d\zeta\left(  1-\frac{\zeta^{2}-a^{2}}
{\sqrt{\left(\zeta^{2}\!+\!a^{2}\!+\!1\right)^{2}\!-\!4a^{2}\zeta^{2}}}\right).
\label{Ga-Zeta}%
\end{equation}
This integral can be transformed to the elliptic form by the substitution
\[
s=\frac{2\sqrt{a^{2}+1}}{\zeta+\left(  a^{2}+1\right)  /\zeta}.
\]
In this case the integration in Eq.\ (\ref{Ga-Zeta}) splits into two segments:
(i) the interval of $\zeta$ between $0$ and $\zeta_{0}=\sqrt{a^{2}+1}$ corresponds
to variation of $s$ from $0$ to $1$ with $\zeta=\frac{1}{s}\sqrt{a^{2}
+1}\left(  1-\sqrt{1-s^{2}}\right)$ and (ii) the interval of $\zeta$ between
$\zeta_{0}$ and $\infty$ corresponds to variation of $s$ from $1$ to $0$ with
$\zeta=\frac{1}{s}\sqrt{a^{2}+1}\left(  1+\sqrt{1-s^{2}}\right)  $. The
substitution transforms $G(a)$ to the following form
\begin{align*}
G(a)  &  =2\sqrt{a^{2}+1}\int\limits_{0}^{1}\frac{ds}{s^{2}}\sum_{\delta=\pm1}
\delta\left(  1+\frac{\delta}{\sqrt{1-s^{2}}}\right) \\
\times &  \left(  1-\frac{1-m_{a}+\delta\left(  1+m_{a}\right)  \sqrt{1-s^{2}
}}{2\sqrt{1-m_{a}s^{2}} }\right) \\
=2  &  \sqrt{a^{2}\!+\!1}\int\limits_{0}^{1}\frac{ds}{\sqrt{1\!-\!s^{2}}
}\left(  2\frac{1\!-\!\sqrt{1\!-\!m_{a}s^{2}}}{s^{2}}+\frac{1-m_{a}}
{\sqrt{1-m_{a}s^{2}}}\right)
\end{align*}
with $m_{a}=\frac{a^{2}}{a^{2}+1}$. The first term in the parenthesis can be
reduced to the full elliptic integrals,
\[
K(m)\!=\!\int\limits_{0}^{1}\frac{dt} {\sqrt{1\!-\!t^{2}}\sqrt{1\!-\!mt^{2}}}
\hbox{ and } E(m)\!=\!\int\limits_{0}^{1}\frac{\sqrt{1\!-\!mt^{2}} }
{\sqrt{1-t^{2}}}dt,
\]
using integration by parts
\begin{align*}
&  \int_{0}^{1}\frac{1-\sqrt{1-ms^{2}}}{s^{2}\sqrt{1-s^{2}}}ds =-\int_{0}
^{1}\left(  1-\sqrt{1-ms^{2}}\right)  d\frac{\sqrt{1-s^{2}}}{s}\\
&  =m\int_{0}^{1}ds\frac{\sqrt{1-s^{2}}}{\sqrt{1-ms^{2}}}=-\left(  1-m\right)
K(m)+E(m).
\end{align*}
This gives us final result
\begin{equation}
G(a)\!=\!4\sqrt{1\!+\!a^{2}}E\left(  \frac{a^{2}}{1\!+\!a^{2}}\right)  \!
-\!\frac{2} {\sqrt{1\!+\!a^{2}}}K\left(  \frac{a^{2}}{1\!+\!a^{2}}\right)
.\label{Ga-Result}%
\end{equation}
This result together with Eq.\ (\ref{Fs-Ga}) describes behavior of $F_{\sigma
}(b,r)$ for $b\ll\pi/r$. Namely, it describes a crossover between the
low-field quadradic regime, $F(b)\approx(3\pi/32)b^{2}$ for $b\ll1$ and the
linear regime, $F(b)\approx b-\pi$ for $1\ll b\ll\pi/r$. To obtain
presentation valid in the whole field range, one can simply add factor
$1-\exp(-\pi/rb)$ to the integral term in Eq.\ (\ref{Fs-Ga})
\begin{equation}
F_{\sigma}(b,r)\!=\!\int\limits_{0}^{\infty}dx\exp\left(  -x\right)  G\left(
\frac{bx} {4}\right)  \left[  1\!-\!\exp\left(  -\frac{\pi} {rb}\right)
\right]  \! -\!\pi.\label{Fs-GaInt}%
\end{equation}
Transformation back to the function $\mathcal{F}_{\sigma}(h,r)$ using
Eq.\ (\ref{Fs-redefine}) gives Eq.\ (\ref{Fs-Present}) of the main text.

\section{Function $\mathcal{F}_{H}(h,r)$ at $r\ll1$.}
\label{App:FH}

In this appendix we obtain a useful presentation of the function $\mathcal{F}
_{H}(h,r)$ defined by Eq.\ (\ref{FHDef}) for $r\ll1$ following the route
similar to one in the Appendix \ref{App-Fs-small-rApp}. First, we introduce
the new reduced field $b=\pi h/r$ and make variable change $u=v/r$ giving the
following presentation
\begin{align*}
\mathcal{F}_{H}(b,r)\!  &  =\!\frac{1}{\pi}\int\limits_{0}^{\infty}
dz\exp\left(  -z\right)  \int\limits_{0}^{\infty}\!dvv\\
\times &  \left\{  \exp\left[  -\frac{\arctan\left[  \left(  v\!+\!bz\right)
/r\right]  \!-\!\arctan\left(  v/r\right)  }{rb}\right]  \right. \\
&  -\left.  \!\!\exp\left[  -\frac{\arctan\left(  v/r\right)  \!-\!\arctan
\left[  \left(  v\!-\!bz\right)  /r\right]  }{rb}\right]  \right\}  .
\end{align*}
In the case $r\ll1,\pi/b$ we can use the asymptotics $\arctan x\approx\pm
\pi/2\mp1/x$ for $x\rightarrow\pm\infty$ in the most part of the integration
domain . We note that the region $0<v<bz$ gives negligible contribution to the
second-term integration because it contains exponentially small factor
$\exp(-\pi/rb)$. Therefore, we can approximate $F_{H}(b,r)$ as
\begin{align*}
\mathcal{F}_{H}(b,r)  &  \approx\frac{1}{\pi}\int\limits_{0}^{\infty}
dz\exp\left(  -z\right)  \left[  \int\limits_{0}^{\infty}vdv\exp\left(
-\frac{z}{v\left(  v\!+\!bz\right)  }\right)  \right. \\
&  -\left.  \int\limits_{bz}^{\infty}vdv\exp\left(  -\frac{z}{v\left(
v-bz\right)  }\right)  \right].
\end{align*}
We see that the dependence on $r$ droped out in this presentation. Shifting
the integration in the second term, $v\rightarrow bz+v$, we derive the
following presentation
\[
\mathcal{F}_{H}(b)\approx\frac{b}{\pi}\int\limits_{0}^{\infty}dv\left[
1\!-\!\int\limits_{0}^{\infty}dzz\exp\left(  -z\!-\frac{z}{v\left(
v\!+\!bz\right)  }\right)  \right]  +\frac{b^{2}}{\pi}.
\]
To remove complicated expression in the exponent, we make the variable change
\begin{align*}
&  x=z+\frac{z}{v^{2}+bzv},\\
z  &  =\frac{xvb-v^{2}-1+\sqrt{\left(  v^{2}-1+xbv\right)  ^{2}+4v^{2}}}{2vb},
\end{align*}
which leads to the following presentation
\begin{equation}
\mathcal{F}_{H}(b)=\frac{b}{\pi}\int_{0}^{\infty}dxx\exp\left(  -x\right)
G_{H}\left(  \frac{bx}{4}\right)  +\frac{b^{2}}{\pi}\label{FHPresApp}%
\end{equation}
with
\begin{align*}
G_{H}(a)  &  =\frac{1}{2}\int_{0}^{\infty}dv\left[  1-\frac{\left(
v^{2}-1+4av\right)  }{\sqrt{\left(  v^{2}-1+4av\right)  ^{2}+4v^{2}}}\right.
\\
&  +\frac{2}{\sqrt{\left(  v^{2}-1+4av\right)  ^{2}+4v^{2}}}\\
&  +\left.  \frac{1}{4av}\left(  1-\frac{\left(  v^{2}+1+4av\right)  }
{\sqrt{\left(  v^{2}-1+4av\right)  ^{2}+4v^{2}}}\right)  \right],
\end{align*}
To transform this function, we first introduce new variable $\zeta$ as
\begin{align*}
&  2\zeta=v-1/v+2a,\\
v  &  =\left(  \zeta-a\right)  +\sqrt{\left(  \zeta-a\right)  ^{2}+1},
\end{align*}
leading to
\begin{align*}
G_{H}(a)\!  &  =\int_{0}^{\infty}d\zeta\left(  1-\frac{\zeta^{2}-a^{2}}
{\sqrt{\left(  \zeta^{2}+a^{2}+1\right)  ^{2}-4\zeta^{2}a^{2}}}\right. \\
&  +\left.  \frac{1}{2\sqrt{\left(  \zeta^{2}+a^{2}+1\right)  ^{2}-4\zeta
^{2}a^{2}}}\right)  -2a.
\end{align*}
This integral reduces to the elliptic form with substitution
\begin{align*}
s  &  =\frac{2\sqrt{a^{2}+1}}{\zeta+\frac{a^{2}+1}{\zeta}},\\
\zeta &  =\frac{\sqrt{a^{2}+1}}{s}\left(  1\pm\sqrt{1-s^{2}}\right)  .
\end{align*}
The $\zeta$ integrations splits into two domains
\begin{align*}
0  &  <\zeta<\zeta_{0}\rightarrow\zeta=\frac{1}{s}\sqrt{a^{2}+1}\left(
1-\sqrt{1-s^{2}}\right)  ,0<s<1\\
\zeta_{0}  &  <\zeta<\infty\rightarrow\zeta=\frac{1}{s}\sqrt{a^{2}+1}\left(
1+\sqrt{1-s^{2}}\right)  ,1>s>0
\end{align*}
with $\zeta_{0}=\sqrt{a^{2}+1}$ and integral for $G_{H}(a)$ becomes
\begin{align*}
&  G_{H}(a)=\sqrt{a^{2}+1}\int\limits_{0}^{1}\frac{ds}{s^{2}}\sum_{\delta
=\pm1}\delta\left(  1+\frac{\delta}{\sqrt{1-s^{2}}}\right) \\
&  \times\left(  1-\frac{1-m_{a}}{4\sqrt{1-m_{a}s^{2}}}-\delta\frac{\left(
3+m_{a}\right)  \sqrt{1-s^{2}}}{4\sqrt{1-m_{a}s^{2}}}\right)  -2a\\
&  =\!2\sqrt{a^{2}\!+\!1}\!\int\limits_{0}^{1}\!\frac{ds}{s^{2}}\frac
{4\sqrt{1\!-\!m_{a}s^{2}}\!-\!1\!+\!m_{a}-\left(  3\!+\!m_{a}\right)  \left(
1\!-\!s^{2}\right)  }{4\sqrt{1-s^{2}}\sqrt{1-m_{a}s^{2}}}\\
&  -2a
\end{align*}
with $m_{a}=a^{2}/(a^{2}+1)$. This integral can be expressed via the full
elliptic integrals $E(m)$ and $K(m)$ as
\begin{equation}
G_{H}(a)=2\sqrt{a^{2}\!+\!1}\left[  E(m_{a})-\frac{1\!-\!m_{a} } {4}
K(m_{a})\right]  \! -\!2a.\label{GHEll}%
\end{equation}
Using asymptotics
\[
G_{H}(a)\approx\frac{\ln\left(  4a\right)  +1}{2a}\text{ for }a\gg1,
\]
we obtain more accurate high-field asymptotics of $\mathcal{F}_{H}(b)$
\[
\mathcal{F}_{H}(b)\approx\frac{b^{2}}{\pi}+\frac{2}{\pi}\left(  \ln
b-\gamma_{E}+1\right)  \text{ for }b\gg1.
\]
Finally, to extend the presentation (\ref{FHPresApp}) to the whole range of
fields including $b>\pi/r$, it is sufficient to add the factor $1-\exp
(-1/h)=1-\exp(-\pi/br)$ to the last term, i.e.,
\begin{align}
\mathcal{F}_{H}(b)  &  \approx\frac{b}{\pi}\int_{0}^{\infty}dxx\exp\left(
-x\right)  G_{H}\left(  \frac{bx}{4}\right) \nonumber\\
&  +\frac{b^{2}}{\pi}\left[  1-\exp\left(  -\frac{\pi}{br}\right)  \right]  .
\end{align}
Returning back to $h=rb/\pi$, we obtain Eq.\ (\ref{FHEll}) of the main text.

\bibliography{MagTranspSDWmetals}

\begin{thebibliography}{39}%
\makeatletter
\providecommand \@ifxundefined [1]{%
 \@ifx{#1\undefined}
}%
\providecommand \@ifnum [1]{%
 \ifnum #1\expandafter \@firstoftwo
 \else \expandafter \@secondoftwo
 \fi
}%
\providecommand \@ifx [1]{%
 \ifx #1\expandafter \@firstoftwo
 \else \expandafter \@secondoftwo
 \fi
}%
\providecommand \natexlab [1]{#1}%
\providecommand \enquote  [1]{``#1''}%
\providecommand \bibnamefont  [1]{#1}%
\providecommand \bibfnamefont [1]{#1}%
\providecommand \citenamefont [1]{#1}%
\providecommand \href@noop [0]{\@secondoftwo}%
\providecommand \href [0]{\begingroup \@sanitize@url \@href}%
\providecommand \@href[1]{\@@startlink{#1}\@@href}%
\providecommand \@@href[1]{\endgroup#1\@@endlink}%
\providecommand \@sanitize@url [0]{\catcode `\\12\catcode `\$12\catcode
  `\&12\catcode `\#12\catcode `\^12\catcode `\_12\catcode `\%12\relax}%
\providecommand \@@startlink[1]{}%
\providecommand \@@endlink[0]{}%
\providecommand \url  [0]{\begingroup\@sanitize@url \@url }%
\providecommand \@url [1]{\endgroup\@href {#1}{\urlprefix }}%
\providecommand \urlprefix  [0]{URL }%
\providecommand \Eprint [0]{\href }%
\providecommand \doibase [0]{http://dx.doi.org/}%
\providecommand \selectlanguage [0]{\@gobble}%
\providecommand \bibinfo  [0]{\@secondoftwo}%
\providecommand \bibfield  [0]{\@secondoftwo}%
\providecommand \translation [1]{[#1]}%
\providecommand \BibitemOpen [0]{}%
\providecommand \bibitemStop [0]{}%
\providecommand \bibitemNoStop [0]{.\EOS\space}%
\providecommand \EOS [0]{\spacefactor3000\relax}%
\providecommand \BibitemShut  [1]{\csname bibitem#1\endcsname}%
\let\auto@bib@innerbib\@empty
\bibitem [{\citenamefont {Paglione}\ and\ \citenamefont
  {Greene}(2010)}]{PaglioneNatPhys10}%
  \BibitemOpen
  \bibfield  {author} {\bibinfo {author} {\bibfnamefont {J.}~\bibnamefont
  {Paglione}}\ and\ \bibinfo {author} {\bibfnamefont {R.~L.}\ \bibnamefont
  {Greene}},\ }\href {http://dx.doi.org/10.1038/nphys1759} {\bibfield
  {journal} {\bibinfo  {journal} {Nat. Phys.}\ }\textbf {\bibinfo {volume}
  {6}},\ \bibinfo {pages} {645} (\bibinfo {year} {2010})}\BibitemShut {NoStop}%
\bibitem [{\citenamefont {Stewart}(2011)}]{StewartRMP11}%
  \BibitemOpen
  \bibfield  {author} {\bibinfo {author} {\bibfnamefont {G.~R.}\ \bibnamefont
  {Stewart}},\ }\href {\doibase 10.1103/RevModPhys.83.1589} {\bibfield
  {journal} {\bibinfo  {journal} {Rev. Mod. Phys.}\ }\textbf {\bibinfo {volume}
  {83}},\ \bibinfo {pages} {1589} (\bibinfo {year} {2011})}\BibitemShut
  {NoStop}%
\bibitem [{\citenamefont {Shibauchi}\ \emph {et~al.}(2014)\citenamefont
  {Shibauchi}, \citenamefont {Carrington},\ and\ \citenamefont
  {Matsuda}}]{ShibauchiARCMP14}%
  \BibitemOpen
  \bibfield  {author} {\bibinfo {author} {\bibfnamefont {T.}~\bibnamefont
  {Shibauchi}}, \bibinfo {author} {\bibfnamefont {A.}~\bibnamefont
  {Carrington}}, \ and\ \bibinfo {author} {\bibfnamefont {Y.}~\bibnamefont
  {Matsuda}},\ }\href {\doibase 10.1146/annurev-conmatphys-031113-133921}
  {\bibfield  {journal} {\bibinfo  {journal} {Annu. Rev. Condens. Matter
  Phys.}\ }\textbf {\bibinfo {volume} {5}},\ \bibinfo {pages} {113} (\bibinfo
  {year} {2014})}\BibitemShut {NoStop}%
\bibitem [{\citenamefont {Hosono}\ and\ \citenamefont
  {Kuroki}(2015)}]{HosonoPhysC15}%
  \BibitemOpen
  \bibfield  {author} {\bibinfo {author} {\bibfnamefont {H.}~\bibnamefont
  {Hosono}}\ and\ \bibinfo {author} {\bibfnamefont {K.}~\bibnamefont
  {Kuroki}},\ }\href {\doibase http://dx.doi.org/10.1016/j.physc.2015.02.020}
  {\bibfield  {journal} {\bibinfo  {journal} {Physica C}\ }\textbf {\bibinfo
  {volume} {514}},\ \bibinfo {pages} {399 } (\bibinfo {year}
  {2015})}\BibitemShut {NoStop}%
\bibitem [{\citenamefont {Gooch}\ \emph {et~al.}(2009)\citenamefont {Gooch},
  \citenamefont {Lv}, \citenamefont {Lorenz}, \citenamefont {Guloy},\ and\
  \citenamefont {Chu}}]{GoochPhysRevB09}%
  \BibitemOpen
  \bibfield  {author} {\bibinfo {author} {\bibfnamefont {M.}~\bibnamefont
  {Gooch}}, \bibinfo {author} {\bibfnamefont {B.}~\bibnamefont {Lv}}, \bibinfo
  {author} {\bibfnamefont {B.}~\bibnamefont {Lorenz}}, \bibinfo {author}
  {\bibfnamefont {A.~M.}\ \bibnamefont {Guloy}}, \ and\ \bibinfo {author}
  {\bibfnamefont {C.-W.}\ \bibnamefont {Chu}},\ }\href {\doibase
  10.1103/PhysRevB.79.104504} {\bibfield  {journal} {\bibinfo  {journal} {Phys.
  Rev. B}\ }\textbf {\bibinfo {volume} {79}},\ \bibinfo {pages} {104504}
  (\bibinfo {year} {2009})}\BibitemShut {NoStop}%
\bibitem [{\citenamefont {Kasahara}\ \emph {et~al.}(2010)\citenamefont
  {Kasahara}, \citenamefont {Shibauchi}, \citenamefont {Hashimoto},
  \citenamefont {Ikada}, \citenamefont {Tonegawa}, \citenamefont {Okazaki},
  \citenamefont {Shishido}, \citenamefont {Ikeda}, \citenamefont {Takeya},
  \citenamefont {Hirata}, \citenamefont {Terashima},\ and\ \citenamefont
  {Matsuda}}]{KasaharaPhysRevB10}%
  \BibitemOpen
  \bibfield  {author} {\bibinfo {author} {\bibfnamefont {S.}~\bibnamefont
  {Kasahara}}, \bibinfo {author} {\bibfnamefont {T.}~\bibnamefont {Shibauchi}},
  \bibinfo {author} {\bibfnamefont {K.}~\bibnamefont {Hashimoto}}, \bibinfo
  {author} {\bibfnamefont {K.}~\bibnamefont {Ikada}}, \bibinfo {author}
  {\bibfnamefont {S.}~\bibnamefont {Tonegawa}}, \bibinfo {author}
  {\bibfnamefont {R.}~\bibnamefont {Okazaki}}, \bibinfo {author} {\bibfnamefont
  {H.}~\bibnamefont {Shishido}}, \bibinfo {author} {\bibfnamefont
  {H.}~\bibnamefont {Ikeda}}, \bibinfo {author} {\bibfnamefont
  {H.}~\bibnamefont {Takeya}}, \bibinfo {author} {\bibfnamefont
  {K.}~\bibnamefont {Hirata}}, \bibinfo {author} {\bibfnamefont
  {T.}~\bibnamefont {Terashima}}, \ and\ \bibinfo {author} {\bibfnamefont
  {Y.}~\bibnamefont {Matsuda}},\ }\href {\doibase 10.1103/PhysRevB.81.184519}
  {\bibfield  {journal} {\bibinfo  {journal} {Phys. Rev. B}\ }\textbf {\bibinfo
  {volume} {81}},\ \bibinfo {pages} {184519} (\bibinfo {year}
  {2010})}\BibitemShut {NoStop}%
\bibitem [{\citenamefont {Hlubina}\ and\ \citenamefont
  {Rice}(1995)}]{HlubRicePhysRevB94}%
  \BibitemOpen
  \bibfield  {author} {\bibinfo {author} {\bibfnamefont {R.}~\bibnamefont
  {Hlubina}}\ and\ \bibinfo {author} {\bibfnamefont {T.~M.}\ \bibnamefont
  {Rice}},\ }\href {\doibase 10.1103/PhysRevB.51.9253} {\bibfield  {journal}
  {\bibinfo  {journal} {Phys. Rev. B}\ }\textbf {\bibinfo {volume} {51}},\
  \bibinfo {pages} {9253} (\bibinfo {year} {1995})}\BibitemShut {NoStop}%
\bibitem [{\citenamefont {Stojkovic}\ and\ \citenamefont
  {Pines}(1997)}]{StojkovicPinesPhysRevB97}%
  \BibitemOpen
  \bibfield  {author} {\bibinfo {author} {\bibfnamefont {B.~P.}\ \bibnamefont
  {Stojkovic}}\ and\ \bibinfo {author} {\bibfnamefont {D.}~\bibnamefont
  {Pines}},\ }\href {\doibase 10.1103/PhysRevB.55.8576} {\bibfield  {journal}
  {\bibinfo  {journal} {Phys. Rev. B}\ }\textbf {\bibinfo {volume} {55}},\
  \bibinfo {pages} {8576} (\bibinfo {year} {1997})}\BibitemShut {NoStop}%
\bibitem [{\citenamefont {L\"ohneysen}\ \emph {et~al.}(2007)\citenamefont
  {L\"ohneysen}, \citenamefont {Rosch}, \citenamefont {Vojta},\ and\
  \citenamefont {W\"olfle}}]{LohneysenRevModPhys.79.1015}%
  \BibitemOpen
  \bibfield  {author} {\bibinfo {author} {\bibfnamefont {H.~v.}\ \bibnamefont
  {L\"ohneysen}}, \bibinfo {author} {\bibfnamefont {A.}~\bibnamefont {Rosch}},
  \bibinfo {author} {\bibfnamefont {M.}~\bibnamefont {Vojta}}, \ and\ \bibinfo
  {author} {\bibfnamefont {P.}~\bibnamefont {W\"olfle}},\ }\href {\doibase
  10.1103/RevModPhys.79.1015} {\bibfield  {journal} {\bibinfo  {journal} {Rev.
  Mod. Phys.}\ }\textbf {\bibinfo {volume} {79}},\ \bibinfo {pages} {1015}
  (\bibinfo {year} {2007})}\BibitemShut {NoStop}%
\bibitem [{\citenamefont {Kontani}(2008)}]{KontaniRPP08}%
  \BibitemOpen
  \bibfield  {author} {\bibinfo {author} {\bibfnamefont {H.}~\bibnamefont
  {Kontani}},\ }\href {http://stacks.iop.org/0034-4885/71/i=2/a=026501}
  {\bibfield  {journal} {\bibinfo  {journal} {Rep. Prog. Phys.}\ }\textbf
  {\bibinfo {volume} {71}},\ \bibinfo {pages} {026501} (\bibinfo {year}
  {2008})}\BibitemShut {NoStop}%
\bibitem [{\citenamefont {Fernandes}\ \emph {et~al.}(2011)\citenamefont
  {Fernandes}, \citenamefont {Abrahams},\ and\ \citenamefont
  {Schmalian}}]{FernandesPhysRevLett11}%
  \BibitemOpen
  \bibfield  {author} {\bibinfo {author} {\bibfnamefont {R.~M.}\ \bibnamefont
  {Fernandes}}, \bibinfo {author} {\bibfnamefont {E.}~\bibnamefont {Abrahams}},
  \ and\ \bibinfo {author} {\bibfnamefont {J.}~\bibnamefont {Schmalian}},\
  }\href {\doibase 10.1103/PhysRevLett.107.217002} {\bibfield  {journal}
  {\bibinfo  {journal} {Phys. Rev. Lett.}\ }\textbf {\bibinfo {volume} {107}},\
  \bibinfo {pages} {217002} (\bibinfo {year} {2011})}\BibitemShut {NoStop}%
\bibitem [{\citenamefont {Breitkreiz}\ \emph
  {et~al.}(2014{\natexlab{a}})\citenamefont {Breitkreiz}, \citenamefont
  {Brydon},\ and\ \citenamefont {Timm}}]{BreitkreizPhysRevB14}%
  \BibitemOpen
  \bibfield  {author} {\bibinfo {author} {\bibfnamefont {M.}~\bibnamefont
  {Breitkreiz}}, \bibinfo {author} {\bibfnamefont {P.~M.~R.}\ \bibnamefont
  {Brydon}}, \ and\ \bibinfo {author} {\bibfnamefont {C.}~\bibnamefont
  {Timm}},\ }\href {\doibase 10.1103/PhysRevB.89.245106} {\bibfield  {journal}
  {\bibinfo  {journal} {Phys. Rev. B}\ }\textbf {\bibinfo {volume} {89}},\
  \bibinfo {pages} {245106} (\bibinfo {year} {2014}{\natexlab{a}})}\BibitemShut
  {NoStop}%
\bibitem [{\citenamefont {Breitkreiz}\ \emph
  {et~al.}(2014{\natexlab{b}})\citenamefont {Breitkreiz}, \citenamefont
  {Brydon},\ and\ \citenamefont {Timm}}]{BreitkreizAnisPhysRevB14}%
  \BibitemOpen
  \bibfield  {author} {\bibinfo {author} {\bibfnamefont {M.}~\bibnamefont
  {Breitkreiz}}, \bibinfo {author} {\bibfnamefont {P.~M.~R.}\ \bibnamefont
  {Brydon}}, \ and\ \bibinfo {author} {\bibfnamefont {C.}~\bibnamefont
  {Timm}},\ }\href {\doibase 10.1103/PhysRevB.90.121104} {\bibfield  {journal}
  {\bibinfo  {journal} {Phys. Rev. B}\ }\textbf {\bibinfo {volume} {90}},\
  \bibinfo {pages} {121104} (\bibinfo {year} {2014}{\natexlab{b}})}\BibitemShut
  {NoStop}%
\bibitem [{\citenamefont {Blomberg}\ \emph {et~al.}(2013)\citenamefont
  {Blomberg}, \citenamefont {Tanatar}, \citenamefont {Fernandes}, \citenamefont
  {Mazin}, \citenamefont {Shen}, \citenamefont {Wen}, \citenamefont {Johannes},
  \citenamefont {Schmalian},\ and\ \citenamefont {Prozorov}}]{BlombergNComm13}%
  \BibitemOpen
  \bibfield  {author} {\bibinfo {author} {\bibfnamefont {E.~C.}\ \bibnamefont
  {Blomberg}}, \bibinfo {author} {\bibfnamefont {M.~A.}\ \bibnamefont
  {Tanatar}}, \bibinfo {author} {\bibfnamefont {R.~M.}\ \bibnamefont
  {Fernandes}}, \bibinfo {author} {\bibfnamefont {I.~I.}\ \bibnamefont
  {Mazin}}, \bibinfo {author} {\bibfnamefont {B.}~\bibnamefont {Shen}},
  \bibinfo {author} {\bibfnamefont {H.-H.}\ \bibnamefont {Wen}}, \bibinfo
  {author} {\bibfnamefont {M.~D.}\ \bibnamefont {Johannes}}, \bibinfo {author}
  {\bibfnamefont {J.}~\bibnamefont {Schmalian}}, \ and\ \bibinfo {author}
  {\bibfnamefont {R.}~\bibnamefont {Prozorov}},\ }\href
  {http://dx.doi.org/10.1038/ncomms2933} {\bibfield  {journal} {\bibinfo
  {journal} {Nat. Commun.}\ }\textbf {\bibinfo {volume} {4}},\ \bibinfo {pages}
  {1914} (\bibinfo {year} {2013})}\BibitemShut {NoStop}%
\bibitem [{\citenamefont {Breitkreiz}\ \emph {et~al.}(2013)\citenamefont
  {Breitkreiz}, \citenamefont {Brydon},\ and\ \citenamefont
  {Timm}}]{BreitkreizPhysRevB13}%
  \BibitemOpen
  \bibfield  {author} {\bibinfo {author} {\bibfnamefont {M.}~\bibnamefont
  {Breitkreiz}}, \bibinfo {author} {\bibfnamefont {P.~M.~R.}\ \bibnamefont
  {Brydon}}, \ and\ \bibinfo {author} {\bibfnamefont {C.}~\bibnamefont
  {Timm}},\ }\href {\doibase 10.1103/PhysRevB.88.085103} {\bibfield  {journal}
  {\bibinfo  {journal} {Phys. Rev. B}\ }\textbf {\bibinfo {volume} {88}},\
  \bibinfo {pages} {085103} (\bibinfo {year} {2013})}\BibitemShut {NoStop}%
\bibitem [{\citenamefont {Rosch}(2000)}]{RoschPhysRevB00}%
  \BibitemOpen
  \bibfield  {author} {\bibinfo {author} {\bibfnamefont {A.}~\bibnamefont
  {Rosch}},\ }\href {\doibase 10.1103/PhysRevB.62.4945} {\bibfield  {journal}
  {\bibinfo  {journal} {Phys. Rev. B}\ }\textbf {\bibinfo {volume} {62}},\
  \bibinfo {pages} {4945} (\bibinfo {year} {2000})}\BibitemShut {NoStop}%
\bibitem [{\citenamefont {Rullier-Albenque}\ \emph {et~al.}(2009)\citenamefont
  {Rullier-Albenque}, \citenamefont {Colson}, \citenamefont {Forget},\ and\
  \citenamefont {Alloul}}]{RullierAlbenquePhysRevLett09}%
  \BibitemOpen
  \bibfield  {author} {\bibinfo {author} {\bibfnamefont {F.}~\bibnamefont
  {Rullier-Albenque}}, \bibinfo {author} {\bibfnamefont {D.}~\bibnamefont
  {Colson}}, \bibinfo {author} {\bibfnamefont {A.}~\bibnamefont {Forget}}, \
  and\ \bibinfo {author} {\bibfnamefont {H.}~\bibnamefont {Alloul}},\ }\href
  {\doibase 10.1103/PhysRevLett.103.057001} {\bibfield  {journal} {\bibinfo
  {journal} {Phys. Rev. Lett.}\ }\textbf {\bibinfo {volume} {103}},\ \bibinfo
  {pages} {057001} (\bibinfo {year} {2009})}\BibitemShut {NoStop}%
\bibitem [{\citenamefont {Rullier-Albenque}\ \emph {et~al.}(2010)\citenamefont
  {Rullier-Albenque}, \citenamefont {Colson}, \citenamefont {Forget},
  \citenamefont {Thu\'ery},\ and\ \citenamefont
  {Poissonnet}}]{RullierAlbenquePhysRevB10}%
  \BibitemOpen
  \bibfield  {author} {\bibinfo {author} {\bibfnamefont {F.}~\bibnamefont
  {Rullier-Albenque}}, \bibinfo {author} {\bibfnamefont {D.}~\bibnamefont
  {Colson}}, \bibinfo {author} {\bibfnamefont {A.}~\bibnamefont {Forget}},
  \bibinfo {author} {\bibfnamefont {P.}~\bibnamefont {Thu\'ery}}, \ and\
  \bibinfo {author} {\bibfnamefont {S.}~\bibnamefont {Poissonnet}},\ }\href
  {\doibase 10.1103/PhysRevB.81.224503} {\bibfield  {journal} {\bibinfo
  {journal} {Phys. Rev. B}\ }\textbf {\bibinfo {volume} {81}},\ \bibinfo
  {pages} {224503} (\bibinfo {year} {2010})}\BibitemShut {NoStop}%
\bibitem [{\citenamefont {Tsukada}\ \emph {et~al.}(2010)\citenamefont
  {Tsukada}, \citenamefont {Hanawa}, \citenamefont {Komiya}, \citenamefont
  {Akiike}, \citenamefont {Tanaka}, \citenamefont {Imai},\ and\ \citenamefont
  {Maeda}}]{TsukadaPhysRevB.81.054515}%
  \BibitemOpen
  \bibfield  {author} {\bibinfo {author} {\bibfnamefont {I.}~\bibnamefont
  {Tsukada}}, \bibinfo {author} {\bibfnamefont {M.}~\bibnamefont {Hanawa}},
  \bibinfo {author} {\bibfnamefont {S.}~\bibnamefont {Komiya}}, \bibinfo
  {author} {\bibfnamefont {T.}~\bibnamefont {Akiike}}, \bibinfo {author}
  {\bibfnamefont {R.}~\bibnamefont {Tanaka}}, \bibinfo {author} {\bibfnamefont
  {Y.}~\bibnamefont {Imai}}, \ and\ \bibinfo {author} {\bibfnamefont
  {A.}~\bibnamefont {Maeda}},\ }\href {\doibase 10.1103/PhysRevB.81.054515}
  {\bibfield  {journal} {\bibinfo  {journal} {Phys. Rev. B}\ }\textbf {\bibinfo
  {volume} {81}},\ \bibinfo {pages} {054515} (\bibinfo {year}
  {2010})}\BibitemShut {NoStop}%
\bibitem [{\citenamefont {Shen}\ \emph {et~al.}(2011)\citenamefont {Shen},
  \citenamefont {Yang}, \citenamefont {Wang}, \citenamefont {Han},
  \citenamefont {Zeng}, \citenamefont {Shan}, \citenamefont {Ren},\ and\
  \citenamefont {Wen}}]{ShenPhysRevB11}%
  \BibitemOpen
  \bibfield  {author} {\bibinfo {author} {\bibfnamefont {B.}~\bibnamefont
  {Shen}}, \bibinfo {author} {\bibfnamefont {H.}~\bibnamefont {Yang}}, \bibinfo
  {author} {\bibfnamefont {Z.-S.}\ \bibnamefont {Wang}}, \bibinfo {author}
  {\bibfnamefont {F.}~\bibnamefont {Han}}, \bibinfo {author} {\bibfnamefont
  {B.}~\bibnamefont {Zeng}}, \bibinfo {author} {\bibfnamefont {L.}~\bibnamefont
  {Shan}}, \bibinfo {author} {\bibfnamefont {C.}~\bibnamefont {Ren}}, \ and\
  \bibinfo {author} {\bibfnamefont {H.-H.}\ \bibnamefont {Wen}},\ }\href
  {\doibase 10.1103/PhysRevB.84.184512} {\bibfield  {journal} {\bibinfo
  {journal} {Phys. Rev. B}\ }\textbf {\bibinfo {volume} {84}},\ \bibinfo
  {pages} {184512} (\bibinfo {year} {2011})}\BibitemShut {NoStop}%
\bibitem [{\citenamefont {Ohgushi}\ and\ \citenamefont
  {Kiuchi}(2012)}]{OhgushiPhysRevB12}%
  \BibitemOpen
  \bibfield  {author} {\bibinfo {author} {\bibfnamefont {K.}~\bibnamefont
  {Ohgushi}}\ and\ \bibinfo {author} {\bibfnamefont {Y.}~\bibnamefont
  {Kiuchi}},\ }\href {\doibase 10.1103/PhysRevB.85.064522} {\bibfield
  {journal} {\bibinfo  {journal} {Phys. Rev. B}\ }\textbf {\bibinfo {volume}
  {85}},\ \bibinfo {pages} {064522} (\bibinfo {year} {2012})}\BibitemShut
  {NoStop}%
\bibitem [{\citenamefont {Eom}\ \emph {et~al.}(2012)\citenamefont {Eom},
  \citenamefont {Na}, \citenamefont {Hoch}, \citenamefont {Kremer},\ and\
  \citenamefont {Kim}}]{EomPhysRevB12}%
  \BibitemOpen
  \bibfield  {author} {\bibinfo {author} {\bibfnamefont {M.~J.}\ \bibnamefont
  {Eom}}, \bibinfo {author} {\bibfnamefont {S.~W.}\ \bibnamefont {Na}},
  \bibinfo {author} {\bibfnamefont {C.}~\bibnamefont {Hoch}}, \bibinfo {author}
  {\bibfnamefont {R.~K.}\ \bibnamefont {Kremer}}, \ and\ \bibinfo {author}
  {\bibfnamefont {J.~S.}\ \bibnamefont {Kim}},\ }\href {\doibase
  10.1103/PhysRevB.85.024536} {\bibfield  {journal} {\bibinfo  {journal} {Phys.
  Rev. B}\ }\textbf {\bibinfo {volume} {85}},\ \bibinfo {pages} {024536}
  (\bibinfo {year} {2012})}\BibitemShut {NoStop}%
\bibitem [{\citenamefont {Sun}\ \emph {et~al.}(2014)\citenamefont {Sun},
  \citenamefont {Taen}, \citenamefont {Yamada}, \citenamefont {Pyon},
  \citenamefont {Nishizaki}, \citenamefont {Shi},\ and\ \citenamefont
  {Tamegai}}]{SunPhysRevB14}%
  \BibitemOpen
  \bibfield  {author} {\bibinfo {author} {\bibfnamefont {Y.}~\bibnamefont
  {Sun}}, \bibinfo {author} {\bibfnamefont {T.}~\bibnamefont {Taen}}, \bibinfo
  {author} {\bibfnamefont {T.}~\bibnamefont {Yamada}}, \bibinfo {author}
  {\bibfnamefont {S.}~\bibnamefont {Pyon}}, \bibinfo {author} {\bibfnamefont
  {T.}~\bibnamefont {Nishizaki}}, \bibinfo {author} {\bibfnamefont
  {Z.}~\bibnamefont {Shi}}, \ and\ \bibinfo {author} {\bibfnamefont
  {T.}~\bibnamefont {Tamegai}},\ }\href {\doibase 10.1103/PhysRevB.89.144512}
  {\bibfield  {journal} {\bibinfo  {journal} {Phys. Rev. B}\ }\textbf {\bibinfo
  {volume} {89}},\ \bibinfo {pages} {144512} (\bibinfo {year}
  {2014})}\BibitemShut {NoStop}%
\bibitem [{\citenamefont {Li}\ \emph {et~al.}(2014)\citenamefont {Li},
  \citenamefont {Yuan}, \citenamefont {Ji}, \citenamefont {Zhang},
  \citenamefont {Ge}, \citenamefont {Feng}, \citenamefont {Yuan}, \citenamefont
  {Hatano}, \citenamefont {Hu}, \citenamefont {Jin}, \citenamefont {Schwarz},
  \citenamefont {Kleiner}, \citenamefont {Koelle}, \citenamefont {Yamaura},
  \citenamefont {Wang}, \citenamefont {Wu}, \citenamefont {Takayama-Muromachi},
  \citenamefont {Vanacken},\ and\ \citenamefont
  {Moshchalkov}}]{LiPhysRevB.90.024512}%
  \BibitemOpen
  \bibfield  {author} {\bibinfo {author} {\bibfnamefont {J.}~\bibnamefont
  {Li}}, \bibinfo {author} {\bibfnamefont {J.}~\bibnamefont {Yuan}}, \bibinfo
  {author} {\bibfnamefont {M.}~\bibnamefont {Ji}}, \bibinfo {author}
  {\bibfnamefont {G.}~\bibnamefont {Zhang}}, \bibinfo {author} {\bibfnamefont
  {J.-Y.}\ \bibnamefont {Ge}}, \bibinfo {author} {\bibfnamefont {H.-L.}\
  \bibnamefont {Feng}}, \bibinfo {author} {\bibfnamefont {Y.-H.}\ \bibnamefont
  {Yuan}}, \bibinfo {author} {\bibfnamefont {T.}~\bibnamefont {Hatano}},
  \bibinfo {author} {\bibfnamefont {W.}~\bibnamefont {Hu}}, \bibinfo {author}
  {\bibfnamefont {K.}~\bibnamefont {Jin}}, \bibinfo {author} {\bibfnamefont
  {T.}~\bibnamefont {Schwarz}}, \bibinfo {author} {\bibfnamefont
  {R.}~\bibnamefont {Kleiner}}, \bibinfo {author} {\bibfnamefont
  {D.}~\bibnamefont {Koelle}}, \bibinfo {author} {\bibfnamefont
  {K.}~\bibnamefont {Yamaura}}, \bibinfo {author} {\bibfnamefont {H.-B.}\
  \bibnamefont {Wang}}, \bibinfo {author} {\bibfnamefont {P.-H.}\ \bibnamefont
  {Wu}}, \bibinfo {author} {\bibfnamefont {E.}~\bibnamefont
  {Takayama-Muromachi}}, \bibinfo {author} {\bibfnamefont {J.}~\bibnamefont
  {Vanacken}}, \ and\ \bibinfo {author} {\bibfnamefont {V.~V.}\ \bibnamefont
  {Moshchalkov}},\ }\href {\doibase 10.1103/PhysRevB.90.024512} {\bibfield
  {journal} {\bibinfo  {journal} {Phys. Rev. B}\ }\textbf {\bibinfo {volume}
  {90}},\ \bibinfo {pages} {024512} (\bibinfo {year} {2014})}\BibitemShut
  {NoStop}%
\bibitem [{\citenamefont {Analytis}\ \emph {et~al.}(2014)\citenamefont
  {Analytis}, \citenamefont {Kuo}, \citenamefont {McDonald}, \citenamefont
  {Wartenbe}, \citenamefont {Rourke}, \citenamefont {Hussey},\ and\
  \citenamefont {Fisher}}]{AnalytisNPhys14}%
  \BibitemOpen
  \bibfield  {author} {\bibinfo {author} {\bibfnamefont {J.~G.}\ \bibnamefont
  {Analytis}}, \bibinfo {author} {\bibfnamefont {H.-H.}\ \bibnamefont {Kuo}},
  \bibinfo {author} {\bibfnamefont {R.~D.}\ \bibnamefont {McDonald}}, \bibinfo
  {author} {\bibfnamefont {M.}~\bibnamefont {Wartenbe}}, \bibinfo {author}
  {\bibfnamefont {P.~M.~C.}\ \bibnamefont {Rourke}}, \bibinfo {author}
  {\bibfnamefont {N.~E.}\ \bibnamefont {Hussey}}, \ and\ \bibinfo {author}
  {\bibfnamefont {I.~R.}\ \bibnamefont {Fisher}},\ }\href {\doibase
  10.1038/NPHYS2869} {\bibfield  {journal} {\bibinfo  {journal} {Nat. Phys.}\
  }\textbf {\bibinfo {volume} {10}},\ \bibinfo {pages} {194} (\bibinfo {year}
  {2014})}\BibitemShut {NoStop}%
\bibitem [{\citenamefont {Moseley}\ \emph {et~al.}(2015)\citenamefont
  {Moseley}, \citenamefont {Yates}, \citenamefont {Peng}, \citenamefont
  {Mandrus}, \citenamefont {Sefat}, \citenamefont {Branford},\ and\
  \citenamefont {Cohen}}]{MoseleyPhysRevB15}%
  \BibitemOpen
  \bibfield  {author} {\bibinfo {author} {\bibfnamefont {D.~A.}\ \bibnamefont
  {Moseley}}, \bibinfo {author} {\bibfnamefont {K.~A.}\ \bibnamefont {Yates}},
  \bibinfo {author} {\bibfnamefont {N.}~\bibnamefont {Peng}}, \bibinfo {author}
  {\bibfnamefont {D.}~\bibnamefont {Mandrus}}, \bibinfo {author} {\bibfnamefont
  {A.~S.}\ \bibnamefont {Sefat}}, \bibinfo {author} {\bibfnamefont {W.~R.}\
  \bibnamefont {Branford}}, \ and\ \bibinfo {author} {\bibfnamefont {L.~F.}\
  \bibnamefont {Cohen}},\ }\href {\doibase 10.1103/PhysRevB.91.054512}
  {\bibfield  {journal} {\bibinfo  {journal} {Phys. Rev. B}\ }\textbf {\bibinfo
  {volume} {91}},\ \bibinfo {pages} {054512} (\bibinfo {year}
  {2015})}\BibitemShut {NoStop}%
\bibitem [{\citenamefont {Jia}\ \emph {et~al.}()\citenamefont {Jia},
  \citenamefont {Welp}, \citenamefont {Marcenat},\ and\ \citenamefont
  {Klein}}]{WelpUnpubl}%
  \BibitemOpen
  \bibfield  {author} {\bibinfo {author} {\bibfnamefont {Y.}~\bibnamefont
  {Jia}}, \bibinfo {author} {\bibfnamefont {U.}~\bibnamefont {Welp}}, \bibinfo
  {author} {\bibfnamefont {C.}~\bibnamefont {Marcenat}}, \ and\ \bibinfo
  {author} {\bibfnamefont {T.}~\bibnamefont {Klein}},\ }\href@noop {}
  {}\bibinfo {note} {Unpublished}\BibitemShut {NoStop}%
\bibitem [{\citenamefont {Watson}\ \emph {et~al.}(2015)\citenamefont {Watson},
  \citenamefont {Yamashita}, \citenamefont {Kasahara}, \citenamefont {Knafo},
  \citenamefont {Nardone}, \citenamefont {B\'eard}, \citenamefont {Hardy},
  \citenamefont {McCollam}, \citenamefont {Narayanan}, \citenamefont {Blake},
  \citenamefont {Wolf}, \citenamefont {Haghighirad}, \citenamefont {Meingast},
  \citenamefont {Schofield}, \citenamefont {v.~L\"ohneysen}, \citenamefont
  {Matsuda}, \citenamefont {Coldea},\ and\ \citenamefont
  {Shibauchi}}]{WatsonPhysRevLett15}%
  \BibitemOpen
  \bibfield  {author} {\bibinfo {author} {\bibfnamefont {M.~D.}\ \bibnamefont
  {Watson}}, \bibinfo {author} {\bibfnamefont {T.}~\bibnamefont {Yamashita}},
  \bibinfo {author} {\bibfnamefont {S.}~\bibnamefont {Kasahara}}, \bibinfo
  {author} {\bibfnamefont {W.}~\bibnamefont {Knafo}}, \bibinfo {author}
  {\bibfnamefont {M.}~\bibnamefont {Nardone}}, \bibinfo {author} {\bibfnamefont
  {J.}~\bibnamefont {B\'eard}}, \bibinfo {author} {\bibfnamefont
  {F.}~\bibnamefont {Hardy}}, \bibinfo {author} {\bibfnamefont
  {A.}~\bibnamefont {McCollam}}, \bibinfo {author} {\bibfnamefont
  {A.}~\bibnamefont {Narayanan}}, \bibinfo {author} {\bibfnamefont {S.~F.}\
  \bibnamefont {Blake}}, \bibinfo {author} {\bibfnamefont {T.}~\bibnamefont
  {Wolf}}, \bibinfo {author} {\bibfnamefont {A.~A.}\ \bibnamefont
  {Haghighirad}}, \bibinfo {author} {\bibfnamefont {C.}~\bibnamefont
  {Meingast}}, \bibinfo {author} {\bibfnamefont {A.~J.}\ \bibnamefont
  {Schofield}}, \bibinfo {author} {\bibfnamefont {H.}~\bibnamefont
  {v.~L\"ohneysen}}, \bibinfo {author} {\bibfnamefont {Y.}~\bibnamefont
  {Matsuda}}, \bibinfo {author} {\bibfnamefont {A.~I.}\ \bibnamefont {Coldea}},
  \ and\ \bibinfo {author} {\bibfnamefont {T.}~\bibnamefont {Shibauchi}},\
  }\href {\doibase 10.1103/PhysRevLett.115.027006} {\bibfield  {journal}
  {\bibinfo  {journal} {Phys. Rev. Lett.}\ }\textbf {\bibinfo {volume} {115}},\
  \bibinfo {pages} {027006} (\bibinfo {year} {2015})}\BibitemShut {NoStop}%
\bibitem [{\citenamefont {Fenton}\ and\ \citenamefont
  {Schofield}(2005)}]{FentonPRL05}%
  \BibitemOpen
  \bibfield  {author} {\bibinfo {author} {\bibfnamefont {J.}~\bibnamefont
  {Fenton}}\ and\ \bibinfo {author} {\bibfnamefont {A.~J.}\ \bibnamefont
  {Schofield}},\ }\href {\doibase 10.1103/PhysRevLett.95.247201} {\bibfield
  {journal} {\bibinfo  {journal} {Phys. Rev. Lett.}\ }\textbf {\bibinfo
  {volume} {95}},\ \bibinfo {pages} {247201} (\bibinfo {year}
  {2005})}\BibitemShut {NoStop}%
\bibitem [{\citenamefont {Lin}\ and\ \citenamefont {Millis}(2005)}]{LinPRB05}%
  \BibitemOpen
  \bibfield  {author} {\bibinfo {author} {\bibfnamefont {J.}~\bibnamefont
  {Lin}}\ and\ \bibinfo {author} {\bibfnamefont {A.~J.}\ \bibnamefont
  {Millis}},\ }\href {\doibase 10.1103/PhysRevB.72.214506} {\bibfield
  {journal} {\bibinfo  {journal} {Phys. Rev. B}\ }\textbf {\bibinfo {volume}
  {72}},\ \bibinfo {pages} {214506} (\bibinfo {year} {2005})}\BibitemShut
  {NoStop}%
\bibitem [{\citenamefont {Koshelev}(2013)}]{KoshelevPRB13}%
  \BibitemOpen
  \bibfield  {author} {\bibinfo {author} {\bibfnamefont {A.~E.}\ \bibnamefont
  {Koshelev}},\ }\href {\doibase 10.1103/PhysRevB.88.060412} {\bibfield
  {journal} {\bibinfo  {journal} {Phys. Rev. B}\ }\textbf {\bibinfo {volume}
  {88}},\ \bibinfo {pages} {060412} (\bibinfo {year} {2013})}\BibitemShut
  {NoStop}%
\bibitem [{Note1()}]{Note1}%
  \BibitemOpen
  \bibinfo {note} {We use system of units with $k_{B}=1$ and $\hbar =1$
  throughout the paper.}\BibitemShut {Stop}%
\bibitem [{\citenamefont {Ziman}(1960)}]{ZimanBook}%
  \BibitemOpen
  \bibfield  {author} {\bibinfo {author} {\bibfnamefont {J.}~\bibnamefont
  {Ziman}},\ }\href@noop {} {\emph {\bibinfo {title} {Electrons and Phonons}}}\
  (\bibinfo  {publisher} {Clarendon, Oxford},\ \bibinfo {year}
  {1960})\BibitemShut {NoStop}%
\bibitem [{\citenamefont {Blatt}(1968)}]{BlattBook}%
  \BibitemOpen
  \bibfield  {author} {\bibinfo {author} {\bibfnamefont {F.~J.}\ \bibnamefont
  {Blatt}},\ }\href@noop {} {\emph {\bibinfo {title} {Physics of electronic
  conduction in solids}}}\ (\bibinfo  {publisher} {New York, McGraw-Hill},\
  \bibinfo {year} {1968})\BibitemShut {NoStop}%
\bibitem [{\citenamefont {Taylor}(1963)}]{Taylor200PRSA63}%
  \BibitemOpen
  \bibfield  {author} {\bibinfo {author} {\bibfnamefont {P.~L.}\ \bibnamefont
  {Taylor}},\ }\href {\doibase 10.1098/rspa.1963.0164} {\bibfield  {journal}
  {\bibinfo  {journal} {Proc. Royal Soc. A}\ }\textbf {\bibinfo {volume}
  {275}},\ \bibinfo {pages} {200} (\bibinfo {year} {1963})}\BibitemShut
  {NoStop}%
\bibitem [{Note2()}]{Note2}%
  \BibitemOpen
  \bibinfo {note} {Note that, in contrast to the relaxation-time approximation,
  the distribution function does not vanish in the hot-spot
  region.}\BibitemShut {Stop}%
\bibitem [{Note3()}]{Note3}%
  \BibitemOpen
  \bibinfo {note} {As the second term vanishes away from the hot spot, in its
  derivation we neglected $p_{s}$ dependence of $v_{s,\alpha }$ and replaced
  $v_{s,\alpha }\rightarrow v_{s,\alpha } ^{\protect \mathrm
  {hs}}$.}\BibitemShut {Stop}%
\bibitem [{\citenamefont {Shockley}(1950)}]{ShockleyPhysRev50}%
  \BibitemOpen
  \bibfield  {author} {\bibinfo {author} {\bibfnamefont {W.}~\bibnamefont
  {Shockley}},\ }\href {\doibase 10.1103/PhysRev.79.191.2} {\bibfield
  {journal} {\bibinfo  {journal} {Phys. Rev.}\ }\textbf {\bibinfo {volume}
  {79}},\ \bibinfo {pages} {191} (\bibinfo {year} {1950})}\BibitemShut
  {NoStop}%
\bibitem [{\citenamefont {Abdel-Jawad}\ \emph {et~al.}(2006)\citenamefont
  {Abdel-Jawad}, \citenamefont {Kennett}, \citenamefont {Balicas},
  \citenamefont {Carrington}, \citenamefont {Mackenzie}, \citenamefont
  {McKenzie},\ and\ \citenamefont {Hussey}}]{AbdelJawadNPhys06}%
  \BibitemOpen
  \bibfield  {author} {\bibinfo {author} {\bibfnamefont {M.}~\bibnamefont
  {Abdel-Jawad}}, \bibinfo {author} {\bibfnamefont {M.~P.}\ \bibnamefont
  {Kennett}}, \bibinfo {author} {\bibfnamefont {L.}~\bibnamefont {Balicas}},
  \bibinfo {author} {\bibfnamefont {A.}~\bibnamefont {Carrington}}, \bibinfo
  {author} {\bibfnamefont {A.~P.}\ \bibnamefont {Mackenzie}}, \bibinfo {author}
  {\bibfnamefont {R.~H.}\ \bibnamefont {McKenzie}}, \ and\ \bibinfo {author}
  {\bibfnamefont {N.~E.}\ \bibnamefont {Hussey}},\ }\href {\doibase
  10.1038/nphys449} {\bibfield  {journal} {\bibinfo  {journal} {Nat. Phys.}\
  }\textbf {\bibinfo {volume} {2}},\ \bibinfo {pages} {821} (\bibinfo {year}
  {2006})}\BibitemShut {NoStop}%
\end{thebibliography}%

\end{document}